\documentclass[tikz, 12pt,a4paper]{article}
    \usepackage{jheppub}
    
\usepackage{amsmath,epsfig}
\usepackage{amssymb,amsfonts}
\usepackage{latexsym}
\usepackage[latin1]{inputenc}
\usepackage{subeqnarray}
\usepackage{xcolor,colortbl}

\usepackage{graphicx}
\usepackage{bbm}
\usepackage{epsfig}

\usepackage[]{mdframed}
\usepackage{multirow}

\usepackage{mdframed}
\usepackage{lipsum} % for creating dummy text
\usepackage{tcolorbox}

\usepackage{pgfplots}
\usepackage{tikz-cd} 
 
\relax

\def\min{\hskip -0.5mm - \hskip -0.5mm}
\def\plus{\hskip -0.5mm + \hskip -0.5mm}

\def\be{\begin{equation}}
\def\ee{\end{equation}}
\def\bea{\begin{eqnarray}}
\def\eea{\end{eqnarray}}
\newcommand\fverb{\setbox\pippobox=\hbox\bgroup\verb}
\newcommand\fverbdo{\egroup\medskip\noindent%
                        \fbox{\unhbox\pippobox}\ }
\newcommand\fverbit{\egroup\item[\fbox{\unhbox\pippobox}]}

\newcommand{\bear}{\begin{eqnarray}}

\newcommand{\eear}{\end{eqnarray}}

\newcommand{\bsea}{\begin{subeqnarray}}
\newcommand{\esea}{\end{subeqnarray}}
\newbox\pippobox

\def\6{\partial}

\def\a{\alpha}

\def\sq
\def\a{\alpha}

 \title{An unusual BPS equation}
                                         \author[a,b]{Constantin Bachas,}
                                           \author[c]{Lorenzo Bianchi}
                                           \author[a]{and Zhongwu Chen}
                                           \affiliation[a]{Laboratoire de Physique de l'Ecole Normale Sup\'erieure,\\
                                           CNRS, PSL Research University and Sorbonne Universit\'es,\\
                                           24 rue Lhomond, 75005 Paris, France}
                                           \affiliation[b]{Physics Department, National Technical University of Athens,\\  15780 Zografou, Athens, Greece} 
                                           \affiliation[c]{Dipartimento di Fisica, Universit\`a di Torino and INFN - Sezione di Torino,\\
                                           Via P. Giuria 1, 10125 Torino, Italy}
                                            \emailAdd{costas.bachas@phys.ens.fr}
                                           \emailAdd{lorenzo.bianchi@unito.it}
                                                                                      \emailAdd{zhongwu.chen@phys.ens.fr}
                                          
                                           \abstract{We prove a conjectured  relation between the   energy-momentum  
                                           and  the 
                                            displacement norm of  superconformal defects. 
                                           The proof  completes  earlier   results, and shows that supersymmetry identifies  two natural  notions
                                            of brane tension in Anti-de Sitter gravity.  As a byproduct we show that a modification of the energy-momentum tensor
                                            that removes the stress of  static superconformal defects,  
                                            ensures  also
                                            that the  radiation these emit  obeys the Null Energy Condition. 
                                            This sheds new light  on  the  radiation-reaction problem for  moving charges. 
                                            }

\begin{document}
                                           \maketitle

%%%%%%%%%%%%%%%%%%%%%%%%%%%%%%%%

 %%%%%%%TeX, LaTeX, %
%%%%%%%%NesTeX}
%%%%%%%%%\dedicated{Dedicated to\ldots\\if you want.}

% Color definitions
\newcommand{\red}[1]{{\color{red} #1 \color{black}}}
\newcommand{\blue}[1]{{\color{blue} #1 \color{black}}}
\newcommand{\green}[1]{{\color{green} #1 \color{black}}}
\newcommand{\yellow}[1]{{\color{yellow} #1 \color{black}}}

\def\pdfannotlink{\pdfstartlink}
 
%\vfil\eject

%%%%%%%%%%%%%%%%%%%%%%%
%%%%%%%%%%%%%%%%%%%%%%%

%%%%%%%%%%%%%%%%%%%%%%%%%%%%%%%%%%%%%%%%%%%%%%%%%%%%%%%%%%%%% 
\section{Introduction}
%mostly stylistic changes, m_0 tomakeclear this is the bare mass

    Two  important observables related to an external probe, or  defect,  of a Quantum Field Theory are  the energy-momentum stored in its  fields, and
   its resistance to deformations. When  the theory is conformal these  are described by two dimensionless parameters,   $a_T$ and $C_D$  \cite{Billo:2016cpy}.
   Lewkowycz and Maldacena observed that a linear relation between them $(C_D = -18 a_T)$ allowed  
    to reconcile two different calculations of the energy emitted by an accelerating half-BPS quark in  ${\cal N}=4$
   super-Yang-Mills  \cite{Lewkowycz:2013laa}. Based on  this and some other examples,  ref.\,\cite{Bianchi:2019sxz}  conjectured  
  that the  relation 
   \bea\label{conj}
\begin{boxed}  {
  {C_D\over a_T} \, =\, -
  {
2(n-1)(p+2)\Gamma (p+1)
\over  
  n\,\pi^{p-{n/2}}\, \Gamma({p\over 2}+1) \Gamma({n-p\over 2})\,   
  }   
  }
  \end{boxed}
%  \qquad \forall \ \  1\leq p\leq n  \hskip -0.5mm - \hskip -0.5mm 2 
 \eea
holds  for any $p$-dimensional superconformal defect in $n$ spacetime dimensions. We will prove this conjecture in the present paper.
\newpage 
 
   A physical interpretation of \eqref{conj} was given in ref.\,\cite{Bachas:2024nvh} which pointed out  that
   in AdS/CFT 
   there are  two ways to  assign an  invariant   tension
   to a $p$-brane dual of a conformal defect. There is a gravitational tension  $\sigma_{\rm gr} = \gamma_{\rm gr}\vert a_T\vert$\, that extends the notion of   ADM  
  mass,\footnote{For earlier discussions of the gravitational or ADM-like  tension see \cite{Harmark:2004ch} and references therein.
   }
    but also an inertial tension or stiffness $\sigma_{\rm in} = \gamma_{\rm in}C_D$. The coefficients $\gamma_{\rm gr}$ and $\gamma_{\rm in}$ were fixed in \cite{Bachas:2024nvh} by  requiring
    that both  tensions reduce to the parameter of the Nambu-Goto action in the limit of  a classical probe brane coupled to gravity.
 The conjecture \eqref{conj}  turns out to imply that $\sigma_{\rm gr} = \sigma_{\rm in} $,  even when  the brane is  quantized and/or   back-reacts strongly. 
  A BPS equation usually relates the mass or tension of an object to its charge. Here it relates two tensions, whence the  title of our paper.

        Clearly $C_D$ is not  defined when $p=0$   (i.e. for local operators),  and   we will see that  $a_T$ is not defined for $p=n\min 1$ (i.e. for  interfaces or boundaries).
 Thus \eqref{conj} applies to $0<p<n\min 1$. Our proof works for  all  superconformal defects that  preserve the transverse-rotation symmetry $\mathfrak{so}(n\min p)$.
 It is guided by,  but  supersedes  earlier  results for  special cases
    \cite{Bianchi:2018zpb,Bianchi:2019sxz,Bianchi:2017ozk,Bianchi:2018scb,Drukker:2020atp}. We only need to prove  \eqref{conj} for the minimal (mutually-compatible)
   number of  bulk and defect supersymmetries. 
 For each pair $(n,p)$ there is at most one 
  such  minimal embedding of  the defect superalgebra into  the bulk superalgebra. The list is given   in  the  table below.

\definecolor{Gray}{gray}{0.92}  
  
\vskip 6pt
   \begin{table}[tbh]
\centering

\begin{tabular}{ |c|c|c|c| } 
\hline
\rowcolor{Gray} defect & ($n, {\cal N}$)  & superalgebra&  $p$-embedding \\  
\hline \hline
\multirow{3}{3em}{\ \ line} & (3,2) &  $\mathfrak{osp}(2\vert 4; \mathbb{R})$  &  $\mathfrak{su}(1,1\vert 1) \oplus \mathfrak{u}(1)_{\rm c}$ \\ 
%\cline{2-4}
& (4,2) & $\mathfrak{su}(2,2\vert 2)$  & $\mathfrak{osp}(4^*\vert 2)$ \\ 
%\cline{2-4}
& (5,1) &  $F(4;2)$ & $D(2, 1;2;0)\oplus \mathfrak{su}(2)_{\rm c}$ \\ 
\hline \hline
\multirow{3}{3em}{surface} & (4,1) & $\mathfrak{su}(2,2\vert 1)$ & $\mathfrak{su}(1,1\vert 1) \oplus \mathfrak{su}(1,1)_{\rm c}\oplus \mathfrak{u}(1)_{\rm c}$ \\ 
%\cline{2-4}
& (5,1) & $F(4;2)$ & $D(2, 1;2;0)\oplus \mathfrak{so}(2,1)_{\rm c}$               \\ 
%\cline{2-4}
& (6,1) & $\mathfrak{osp}(8^*\vert 2) $ & $\mathfrak{osp}(4^*\vert 2) \oplus    \mathfrak{so}(2,1)_{\rm c} \oplus \mathfrak{so}(3)_{\rm c}$ \\ 
\hline\hline
$p=3$ & (5,1) & $F(4;2)$ & $\mathfrak{osp}(2\vert 4; \mathbb{R}) \oplus \mathfrak{u}(1)_{\rm c}$ \\ 
\hline\hline
$p=4$ & (6,1) & $\mathfrak{osp}(8^*\vert 2) $ & $\mathfrak{su}(2,2\vert 1) \oplus \mathfrak{u}(1)_{\rm c}$ \\ 
\hline
 \end{tabular}
 \vskip 2pt
\caption{\footnotesize The minimal supersymmetric  DCFTs discussed in the paper.  
 The  second  column  gives the smallest ${\cal N}$ that a $n$-dimensional 
 SCFT  must have to admit   $p$-dimensional superconformal defects. 
 The corresponding  bulk superalgebras and maximal $p$-embeddings are given in the third and fourth columns.
 The subscript `c' denotes bosonic subalgebras that commute with the preserved supercharges (for details see section \ref{sec:2}).
 For missing $(p,n)$ pairs  superconformal and rotation-invariant defects do not exist. }
\label{tab1}
\end{table}

%\vskip 1mm

 Our proof relies on  reformulating  \eqref{conj} as a  property of the  two-point function 
 $\langle\hskip -1mm \langle   T^{\,\mu\nu} D^j\rangle\hskip -1mm\rangle$, 
where $D^j$ is the displacement operator. This will  shed  light on 
the role of supersymmetry for defect dynamics.   
We  mentioned already that  \eqref{conj} was  originally motivated by  studies  of the radiation from  an accelerating `quark'
\cite{Mikhailov:2003er,Athanasiou:2010pv,Hubeny:2010bq,Hatta:2011gh,Baier:2011dh,Correa:2012at,Fiol:2012sg,Agon:2014rda,Fiol:2015spa,Fiol:2019woe,Bianchi:2019dlw}. 
What the authors of \cite{Lewkowycz:2013laa} observed  was that in  ${\cal N}$\,=\,$4$
   super-Yang-Mills   one can reconcile  two calculations of the emitted  energy,    proportional  respectively  to 
   $C_D$ and $\vert a_T\vert$, by
     changing  the bulk energy-momentum tensor $T^{\mu\nu} \to \tilde T^{\mu\nu}$.
  They suggested that this works because  $\tilde T^{\mu\nu}$ separates
   radiation  from the quark  self-energy,  but
%correct   but 
   the precise reason
   remained unclear.
  It was later 
 %correct  shown
 pointed out  
   in ref.\,\cite{Fiol:2019woe} 
   that  for a free scalar field such a  modification of $T^{\mu\nu}$
removes  the conformal-improvement term   
   which  is known to violate  the Null Energy Condition (NEC) \cite{Visser:1999de}. 
   
   We will show that the two effects are related.  The key point  is that the BPS equation  \eqref{conj} guarantees  that the
   {\it same} modification
   %correct that 
   which restores the NEC in the radiation flux  also removes the stress stored in the fields of a static defect.
Zero stress  is necessary if one wants to absorb   the leading singularity of $T^{\mu\nu}$  by  renormalizing  the (infinite) bare mass of the defect.
 This  argument   suggests  that supersymmetry may help solve some  time-honoured puzzles of radiation 
 reaction for moving charges in  electrodynamics \cite{Dirac,FR,LL,Teitelboim:1970mw,Poisson:1999tv,Galtsov:2004uqu}. We
 defer this  problem  to   future work.

In two-dimensional CFT the 2-point function  $\langle\hskip -1mm \langle   T^{\,\mu\nu} D^j\rangle\hskip -1mm\rangle$ encodes
 universal aspects of energy transport  across a conformal defect \cite{Meineri:2019ycm}.
Since  $a_T$  is not defined in this case,  the energy transport depends 
 only on $C_D$,  or 
   in the dual AdS$_3$ gravity   on $\sigma_{\rm in}$ \cite{Bachas:2020yxv,Bachas:2021tnp,Baig:2022cnb,Bachas:2022etu}. 
Part of our  motivation in the present work  was  to  extend these  studies  to    higher dimensions.
Let us  finally mention   that  for even-dimensional defects eq.\,\eqref{conj}  relates two anomaly coefficients
   \cite{Graham:1999pm,Bianchi:2015liz,Jensen:2018rxu,Chalabi:2020iie,Chalabi:2021jud}.

%%%%%%%%%%%%%%%%%%%%%%%%%%%%

\subsection{Definitions and outline}
We now define more precisely  the two main  characters in  \eqref{conj}. 
    A  Defect Conformal Field Theory (DCFT) has  in addition to the bulk stress tensor $T^{\mu\nu}$,    a displacement operator 
   $D^\nu$ living  on the defect. It is defined through  the Ward identity of  broken translation invariance
   \bea\label{conserve}
    \partial_\mu T^{\mu\nu}   = \delta_{\rm def}(x) \,   D^\nu \ , 
   \eea
   where $\delta_{\rm def}$ is  the delta function localized on the 
  %correct
   defect.\footnote{When Lorentzian 
   our defects 
   are always timelike. The word  `defect'  is used for both the defect  and the defect's worldvolume.
     The  meaning should be  clear from the context. }
    For a  (hyper)planar defect  
   in  $\mathbb{R}^{(1,n-1)}$  the only non-vanishing components of  $D^\nu$
  are  transverse. 
  %,   $D^j$ with $j\in\{p+1, \cdots , n\}$\,. 

   In the background of such a (hyper)planar  defect  the
  one-point function  $\langle \hskip -0.9mm \langle T^{\mu\nu} \rangle \hskip -0.9mm \rangle $  and the two point-function  
  $\langle \hskip -0.9mm \langle D^j D^k  \rangle \hskip -0.9mm \rangle $  are  fixed   by the unbroken   $\mathfrak{so}(2,p)\oplus \mathfrak{so}(n\min p)$  symmetry modulo
  two free parameters $a_T$ and $C_D$,  \cite{Billo:2016cpy} 
   \begin{subequations}\label{26}
\begin{align}
&\langle\hskip -1mm \langle   T^{\,\alpha\beta}(x) \rangle\hskip -1mm\rangle
\,=\, 
 {a}_{T} 
  \, \bigl(  {\scalebox{0.95}{$n-p-1$}
\over n}\bigr)\, 
{\eta^{\alpha\beta}\over \vert x_\perp\vert^n}
\ , \qquad
\langle\hskip -1mm \langle   T^{\,\alpha j}(x) \rangle\hskip -1mm\rangle\, = \, 0\ ,  \label{26a}
 \\ 
 &\ \quad  \langle\hskip -1mm \langle   T^{\,ij}(x) \rangle\hskip -1mm\rangle\, = \,  a_{T} \,\Bigl[ 
   \,{x^i x^j \over \vert x_\perp\vert^{n+2}}
   \,-\,
 \bigl({  \scalebox{0.95}{$p+1$}\over n }\bigr)
{\delta^{\,ij} \over \vert x_\perp\vert^n} \, 
 \Bigr]\ \ ,  \label{26b}
\end{align} 
\end{subequations}
\vskip -12pt
   \bea\label{27}
  {\rm and} \qquad  \langle \hskip -0.9mm \langle D^j(y)  D^k(0)  \rangle \hskip -0.9mm \rangle  = C_D\,  {\delta^{\,jk} \over \vert y\vert^{2p+2}}\ .  \qquad
  % \langle \hskip -0.9mm \langle T^{\mu\nu}(x) \rangle \hskip -0.9mm \rangle = 
   \eea
% correct2: delta missing
We  here separated parallel and transverse directions, $x^\mu=(x^\alpha, x^j)$ and $y^\mu = (y^\alpha, 0)$, and
denoted   the distance of $x$ from the defect $\vert x_\perp\vert$\,.   
Our notation is described in more detail in section \ref{TDco}. 
  Note that the conservation equation \eqref{conserve} fixes
the normalization of $D^j$, so that $C_D$  is a piece of DCFT data. Note also that 
 the right-hand sides in \eqref{26}
%$\langle \hskip -0.9mm \langle T^{\mu\nu} \rangle \hskip -0.9mm \rangle $ 
vanish  identically when $p=n\min 1$,  which is why
 $a_T$ is  not defined for interfaces and boundaries.

 Unitarity implies $C_D> 0$, while
 positivity of $T^{00}$  implies that  $a_T < 0$.\footnote{\,The spacetime signature is mostly plus. Some of the literature,  including 
 ref.\,\cite{Bianchi:2019sxz}, 
   uses instead of $a_T$ the   parameter $h = -a_T (n$$-$$p$$-$$1) /n$\,.
 %Both $C_D$ and $a_T$ actually vanish in the  extremal case of a topological defect.
 } 
% correct
The two pieces of DCFT data are otherwise unconstrained. They are in general different functions of the marginal couplings and other parameters, such as the ranks of gauge groups.
This is  illustrated by the free-field example of section \ref{sec:abelian}. 
%The BPS relation  \eqref{conj}  only depends on the defect and spacetime dimensions  $p$ and  $n$. 

 % and expressed the conjecture as  \bea\nonumber
%  C_D\, =\, h \,\,   {
%2^{p+1} (n-1)(p+2)\Gamma ({p+1\over 2})\,\pi^{(nn-p)/2}
%\over    (n-p-1)\,\pi^{(p+1)/2}\,   \Gamma({n-p\over 2})  }\ . 
%\eeaThat this is equivalent to \eqref{conj} follows from gamma-function identities.   } 
%%   

\vskip 1pt

 The paper is structured as follows. In section \ref{sec:2}  we explain  table \ref{tab1}. 
For each pair $(p,n)$ we find the minimal
 bulk  supersymmetry ${\cal N}$ that admits rotation-invariant superconformal defects. We do this by examining  all 
   superalgebra embeddings,
  along the lines of  \cite{DHoker:2008wvd, Gutperle:2017nwo, Agmon:2020pde}.
 A subtle point about  real forms of  the 5d ${\cal N}=1$ superalgebra is discussed in appendix \ref{F42}.

In section \ref{TDcor} we introduce the two-point function  $\langle\hskip -1mm \langle   T^{\,\mu\nu} D^j\rangle\hskip -1mm\rangle$, 
which is determined  in terms of $a_T$ and $C_D$ by conformal Ward identities \cite{Billo:2016cpy}. 
The  relation \eqref{conj} takes a particularly simple form when expressed as a relation between  the coefficients of 
various tensor structures in $\langle\hskip -1mm \langle   T^{\,\mu\nu} D^j\rangle\hskip -1mm\rangle$. 
This is the starting point of our proof. 
We put some flesh into these  abstract formulae with  the free-field example of   line defects in the 4d ${\cal N}$ = 2 abelian gauge theory.

Section \ref{sec:4}  explains how the modified stress tensor introduced in \cite{Lewkowycz:2013laa} can be tuned to cancel  either the stress of  the static fields, or
the NEC-violating radiation from an accelerating defect. {\it The BPS equation \eqref{conj} ensures that both are simultaneously cancelled}. 
The conclusion only depends on the structure of $\langle\hskip -1mm \langle   T^{\,\mu\nu} D^j\rangle\hskip -1mm\rangle$, and is valid
for an arbitrary  DCFT. Sections \ref{TDcor} and \ref{sec:4} are autonomous, 
readers only interested in the problem of radiation reaction can skip the rest of the paper. 

 Section \ref{sec:proof} presents the proof of \eqref{conj}. The BPS equation is  formulated as  a 
 special property of $\langle\hskip -1mm \langle   T_{zz} D_z\rangle\hskip -1mm\rangle$,  where $z$  is  a complex transverse direction.
This property  follows then  from supersymmetric  Ward identities that have the  same   form for all $(n,p)$. 
The   superalgebras of table \ref{tab1} are summarized for completeness in appendix  \ref{algebras}. 
 The proof uses the supersymmetry  transformations of the displacement and stress-tensor multiplets, which  are
summarized in appendix \ref{app:C}.

%%%%%%%%%%%%%%%%%
%%%%%%%%%%%%%%%%%%%%%%%%%%%%%%%%%%%%%%%%%%%%%%%%%%%%%%%%%%%%%%%%%%%%%%%%%%%%%

\section{Minimal  superconformal defects}\label{sec:2} 

This section explains the entries of  table \ref{tab1}. 
 We  adapt  to our purposes  and extend the work 
 of  refs.\,\cite{DHoker:2008wvd, Gutperle:2017nwo, Agmon:2020pde} 
that  classified  various  superconformal  defects.

   We are  interested in defects for which both $C_D$ and $a_T$ are defined. This  excludes local operators and domain walls, 
 so  \,  $1\leq p\leq n-2$. 
Since superconformal theories (SCFTs) only exist in $n\leq 6$ dimensions \cite{Nahm:1977tg},  
 we  only need to consider    $n=3,4,5,6$.  The  corresponding   superconformal algebras $\mathfrak{G}_{\rm s}^{(n)}$,   consistent with   the existence of
a  unitary conserved stress-tensor   multiplet,   are well known
  \cite{Minwalla:1997ka}\cite{Cordova:2016emh}. We list them  below for the reader's convenience
\bea\label{listSCFT}
& \hskip 1.2mm  n=3: \ \ 
\mathfrak{osp}({\cal N}\vert 4; \mathbb{R}) 
\ \overset{bos}{\supset} \ 
 \mathfrak{so}(2,3) \oplus \mathfrak{so}({\cal N})_{R}\ , \hskip 5mm {\cal N}= 1, \cdots ,6, 8 \ ; \nonumber \\[2pt]
&\hskip 0.3mm
n=4: \hskip 3.6mm
  \mathfrak{su}(2, 2\vert {\cal N}) \ \overset{bos}{\supset} \  \mathfrak{so}(2,4) \oplus \mathfrak{u}({\cal N})_R\ , \quad \ \  \ \  {\cal N}= 1, 2,3   \ ;\quad 
   \nonumber\\[2pt]
  &\quad \hskip 2mm  \mathfrak{psu}(2, 2\vert {\cal N}) \ \overset{bos}{\supset} \  \mathfrak{so}(2,4) \oplus \mathfrak{su}(4)_R\ , \quad \ \    {\cal N}= 4 \ ;
 \\[2pt]
  & n=5:\ \   
   F(4;2) \ \overset{bos}{\supset} \  \mathfrak{so}(2,5) \oplus \mathfrak{su}(2)_R\ , \quad \hskip 12mm  {\cal N}= 1  \ ; \qquad \hskip 2mm   \nonumber\\[2pt]
    &\hskip -1.2mm  n=6:\ \    \mathfrak{osp}(8^*\vert 2{\cal N}) \ \overset{bos}{\supset} \  \mathfrak{so}(2,6)\oplus \mathfrak{usp}(2 {\cal N})_R\ , \hskip 4mm {\cal N}= 1,2  \ . \quad 
    \nonumber
\eea
As usual,  ${\cal N}$  is  the number of  irreducible  $\mathfrak{so}(1,n\min 1)$ spinor   supercharges.
There is one superalgebra for each pair $({\cal N},n)$. 
Also  given  in \eqref{listSCFT}
are  the  maximal bosonic (even-degree) subalgebras $ \mathfrak{so}(2,n) \oplus \mathfrak{R} $,  where $ \mathfrak{so}(2,n)$ is the conformal group and
$\mathfrak{R}$
is the  compact  R-symmetry.\,\footnote{\,We will use the  special inclusion symbol $\ \overset{bos}{\supset} \ $  for  the bosonic
 subalgebra of a superalgebra.}
 
\vskip 1pt

   A planar superconformal $p$-dimensional defect breaks $\mathfrak{G}_{\rm s}^{(n)} \hookrightarrow \mathfrak{g}_{\rm s} \oplus \mathfrak{g}_{\rm c}$\,,  where 
   $\mathfrak{g}_{\rm s}$ is the  defect superalgebra
   and 
   $\mathfrak{g}_{\rm c}$  
 a    bosonic algebra  that commutes with all preserved  supercharges.
 %\footnote{\,
 Most of the time   $\mathfrak{g}_{\rm c}$ is a compact symmetry that may,  but need not,    be broken by the 
 superconformal defect. 
     But for  surface defects   it may include a chiral  $\mathfrak{so}(2,1)$ half  of the conformal symmetry. 
  %   }

  We  restrict attention  to defects    invariant under both worldvolume and transverse rotations.
 This condition was already implicit 
in  eqs.\,\eqref{26} and \eqref{27}. It  implies 
 that  the displacement operator sits in an   irreducible multiplet  of  the defect algebra  \cite{Agmon:2020pde}. 
It also implies (the last of)  the following chain of  inclusions  
 \bea
\hskip -4mm   \mathfrak{so}(2,n) \hskip -0.6mm \oplus  \hskip -0.5mm \mathfrak{R}
\   {\supset} \   
 \mathfrak{so}(2,p) \hskip -0.6mm\oplus\hskip -0.6mm \mathfrak{so}(n\hskip -0.6mm -\hskip -0.6mm p) \hskip -0.6mm\oplus \hskip -0.5mm  \mathfrak{R}
 \, \supset \,  
  \mathfrak{g}_{\rm s}^{\rm bos} \hskip -0.5mm\oplus\hskip -0.3mm  \mathfrak{g}_{\rm c} 
  \, \supset \,  
  \mathfrak{so}(2,p)\hskip -0.6mm\oplus \hskip -0.6mm\mathfrak{so}(n\hskip -0.6mm -\hskip -0.6mm p)_{\rm def}\,. 
 \eea
 Here $\mathfrak{g}_{\rm s}^{\rm bos}= \mathfrak{so}(2,p)\hskip -0.6mm\oplus \mathfrak{r}$ is the bosonic subalgebra of $\mathfrak{g}_{\rm s}$, 
 with $\mathfrak{r}$ the defect R-symmetry. 
 The 
 subscript  `def'   indicates that if  a simple component of $\mathfrak{so}(n\hskip -0.6mm -\hskip -0.6mm p)$  fits in  $\mathfrak{R}$, 
  the preserved   rotations could  be accompanied   by bulk R-transformations.\footnote{\,More generally, if the bulk theory 
has  a  flavour symmetry, the unbroken  rotations could act non-trivially in flavour space.  
Since both the stress-tensor and the displacement supermultiplets are `flavour blind', 
this  will  not affect our proof.  }  
%We will see that  this only happens for codimension-two defects. 

We call  the  inclusion  of $\mathfrak{g}_{\rm s} \oplus \mathfrak{g}_{\rm c}$ in $ \mathfrak{G}_{\rm s}^{(n)}  $
 a  `$p$-embedding'
if the   non-compact  unbroken symmetry  is $\mathfrak{so}(2,p)$\,.
 A  $p$-embedding is  {\it maximal} if it cannot be  properly included in any other $p$-embedding. 
 Note that a $p$-embedding may be included  in a  higher-$p$  embedding, as befits the symmetry structure 
of  a defect inside a  defect. Such 
composite defects break the worldvolume and/or transverse-rotation symmetry,  and we do not discuss them here.
  
 Our  problem is now  to  classify  $p$-embeddings
 in the bulk superalgebras   \eqref{listSCFT}, for $p\leq n$$-$2\,. 
Fortunately,  maximal  real  subalgebras of  the $\mathfrak{osp}$ and $\mathfrak{su}$ series and of $F(4;2)$  
  have been  classified \cite{DHoker:2008wvd,Gutperle:2017nwo,Agmon:2020pde}, 
and  are  collected conveniently in table 2 of ref.\,\cite{Agmon:2020pde}.\footnote{A bug in this table will be corrected in  section \ref{2in5}.} 
If  there is no maximal 
 $p$-embedding for some $ \mathfrak{G}_{\rm s}^{(n)}  $ in the  list, we conclude that the  corresponding SCFT  does not admit  stand-alone  $p$-dimensional 
 superconformal defects.

 %%%%%%%%%%%%%%%%%%%%%%%%%%%%%%%
 %%%%%%%%%%%%%%%%%%
 \subsection{Line defects} 
We begin   with the  line defects, $p=1$, which were   classified (without assuming rotation symmetry)  in table 12 of
ref.\,\cite{Agmon:2020pde}.

 Inspection of  this  table  shows,  first  of all,   that bulk theories with  minimal,   ${\cal N}=1$,  supersymmetry in 
 $n=3,4$ and $6$ dimensions do not admit  any superconformal  line defects. 
  One can  understand why  using elementary   spinor properties.\footnote{Our argument uses rotation  invariance,  
 but  breaking it does not help, see  \cite{Agmon:2020pde}. 
 }
    Indeed, the ${\cal N}=1$ R-symmetry    is too small  to contain the group 
 of   transverse  rotations,  so the preserved  $\mathfrak{so}(n\hskip -0.6mm -\hskip -0.6mm 1)_{\rm def}$ is   canonically embedded   in the ambient Lorentz group 
 $ \mathfrak{so}(1,n\hskip -0.6mm -\hskip -0.6mm 1)$. 
  It follows that the bulk Poincar\'e supercharges,  $Q $,  must transform as  an irreducible spinor of  both   $\mathfrak{so}(1,n\hskip -0.6mm -\hskip -0.6mm 1)$, 
since supersymmetry is minimal, 
and  
of   $\mathfrak{so}(n\hskip -0.6mm -\hskip -0.6mm 1)_{\rm def}$. 
 But  for $n=3,4$ and $6$ the irreducible spinors of   $\mathfrak{so}(n\hskip -0.6mm -\hskip -0.6mm 1)$ 
  have the same dimension as  those of $\mathfrak{so}(1, n\hskip -0.6mm -\hskip -0.6mm 1)$ (see e.g. appendix B of ref.\,\cite{Polchinski:1998rr}). 
 Since the  defect must  break  some  
 supersymmetries, this proves  that   the above ${\cal N}=1$ SCFTs  have no  rotation-invariant  superconformal line defects.

This simple counting   argument  does not  exclude  line defects   in the six-dimensional ${\cal N}=(2,0)$ theory. 
The    bulk Poincar\'e supercharges transform  in this case in the $(\boldsymbol{4}, \boldsymbol{2},\boldsymbol{1}) \oplus (\boldsymbol{4}, \boldsymbol{1},\boldsymbol{2})$ 
representation of 
$\mathfrak{so}(1,5)\oplus \mathfrak{su}(2) \oplus \mathfrak{su}(2)^\prime$\,, 
where 
$\mathfrak{su}(2) \oplus \mathfrak{su}(2)^\prime\subset  \mathfrak{so}(5)_R\simeq \mathfrak{usp}(4)_R$\,\cite{Polchinski:1998rr}. 
A half-BPS line  defect could,  a priori,   reduce these   to one
irreducible  spinor of $\mathfrak{so}(5)_{\rm def}$, but the rigidity of  superconformal algebras forbids it. 
Indeed, none of the maximal subalgebras of  $\mathfrak{osp}(8^*\vert 4)$ can describe the symmetries   of an odd-dimensional defect \cite{Agmon:2020pde}. 
The best one  can do is to consider maximal 1-embeddings  in some higher-$p$ subalgebra of $ \mathfrak{osp}(8^*\vert 4)$, but this breaks
the $\mathfrak{so}(5)$ rotation symmetry as we have explained. 
% We conclude that  there are no  superconformal rotation-invariant line defects in any six-dimensional theory.

 The case $n=5$ is special because a Weyl-Majorana spinor of 
 $\mathfrak{so}(4)$ has half as many components as an irreducible  spinor of  $\mathfrak{so}(1,4)$\,. 
 Rotation-invariant  
 %\scalebox{0.98}{${1\over 2}$}
  line defects are  in this case allowed. They are described by the maximal  1-embedding
    \bea\label{1in5}
 F(4;2)\ \supset \ D(2,1;2;0) \oplus \mathfrak{su}(2)_{\rm c} \ \overset{bos}{\supset} \  \mathfrak{so}(2,1)\oplus \mathfrak{so}(4)  \oplus \mathfrak{su}(2)_{\rm c} \ . 
 \eea
 Note that the bosonic subalgebra contains  three $\mathfrak{su}(2)$ factors, one  
coming from   the bulk R-symmetry,   $\mathfrak{su}(2)_R$, 
and  the others  from  $\mathfrak{so}(4)   \simeq \mathfrak{su}(2)\oplus \mathfrak{su}(2)^{\,\prime}$. 
 The defect R-symmetry is $\mathfrak{so}(4)  \simeq \mathfrak{su}(2)\oplus \mathfrak{su}(2)_R$\,
while  $\mathfrak{su}(2)_{\rm c} $=$\mathfrak{su}(2)^{\,\prime}$ commutes with the supercharges (see  appendix \ref{F42}). 
Line defects of  the five-dimensional SCFT have been analyzed  in  refs.\,\cite{Assel:2012nf,Dibitetto:2018gtk,Chen:2019qib,Uhlemann:2020bek}\,.

\vskip 1pt

 Superconformal  line defects also exist in   three-  and four-dimensional SCFTs but with  ${\cal N}=2$ supersymmetry. 
  The relevant   maximal 1-embeddings are 
     \bea\label{1in3}
 n=3: \quad \mathfrak{osp}(2\vert  4; \mathbb{R})  \ \supset \ \mathfrak{su}(1,1\vert 1)\oplus \mathfrak{u}(1)_{\rm c}  \ \overset{bos}{\supset} \ 
  \mathfrak{so}(2,1)\oplus \mathfrak{so}(2)_{\rm def} \oplus \mathfrak{u}(1)_{\rm c} \  ; 
\eea \vskip -5mm
\bea
  \hskip -12mm 
n=4: \quad \mathfrak{su}(2, 2\vert  2)  \ \supset \ \mathfrak{osp}(4^*\vert 2)\ \overset{bos}{\supset} \  \mathfrak{so}(2,1)\oplus \mathfrak{so}(3) \oplus \mathfrak{usp}(2) \  . 
 \eea
In three dimensions
the preserved  rotation symmetry  
  is a mixture of space rotations and  R-symmetry transformations, and
 %It is 
% in general non-trivially embedded in  the bulk   $\mathfrak{so}(2)\oplus \mathfrak{u}(1)_R$, while the  
the commuting   $\mathfrak{u}(1)_{\rm c}$ may (but need not) be  broken (for an overview on explicit constructions of 3d line defects see \cite{Drukker:2019bev} and references therein).
In four dimensions, on the other hand, both $\mathfrak{so}(3)$ and  $\mathfrak{usp}(2)_R$ are part of the defect superalgebra that  has no commutant  in  $\mathfrak{G}^{(4)}_{\rm s}$. 
% Many explicit  constructions  of such line defects  are known, see \cite{Agmon:2020pde} for a collection of  references. 
\vskip 1pt

  Extended ${\cal N}>2$ supersymmetry allows a large  variety of   BPS line defects. These  were  classified in ref.\,\cite{Agmon:2020pde} where the reader can also find  references to explicit constructions. 
  The  defect  superalgebras  in three dimensions are 
  $\mathfrak{su}(1,1\vert m)$ with $m=1,3,4$\,, or $\mathfrak{psu}(1,1\vert 2)$,
  and in four dimensions $\mathfrak{osp}(4^*\vert 2m)$ with $m=1,2$ (modulo  bosonic commutants  $\mathfrak{g}_{\rm c}$). 
 In all cases the unbroken symmetry contains the minimal defect 
 superalgebras
    $\mathfrak{su}(1,1\vert 1)$  for $n=3$ and   $\mathfrak{osp}(4^*\vert 2)$ for $n=4$.  
These are   sufficient,  as we will see,  for   proving   the conjecture   \eqref{conj}.

      To summarize our discussion up to here, minimal  superconformal  line defects exist   in $n=3,4$ and $5$-dimensional SCFTs with 
     ${\cal N}=2,2$ and  $1$ supersymmetries. The corresponding defect superalgebras are $\mathfrak{su}(1,1\vert 1)$, $\mathfrak{osp}(4^*\vert 2)$ and $D(2,1;2;0)$, and  
    the  unbroken   supercharges  transform as spinors of $\mathfrak{so}(n\hskip -0.6mm -\hskip -0.6mm 1)$\,. 
     Simple counting shows that   these  defects are half-BPS,  and that there exist  no other, less supersymmetric line defects.

 %%%%%%%%%%%%%%%%%%%%%%%%%%%%%%%
 %%%%%%%%%%%%%%%%%%
 \subsection{Surface defects} \label{2in5}

 We move  next  to $p=2$ and $n=4,5, 6$.  The  conformal group of a surface defect factorizes  into a left- and a right-moving piece, 
 $ \mathfrak{so}(2,2)\simeq \mathfrak{so}(2,1)_+\oplus \mathfrak{so}(2,1)_-$\,. Each   can  be extended separately  to a one-dimensional  superconformal algebra. 
  %This is indeed the case   in $n=4$ and $n=6$ dimensions.  
 
  Consider  first the  ${\cal N}=1$   four-dimensional SCFTs,  whose Lie superalgebra is $\mathfrak{su}(2,2\vert 1)\supset
 \mathfrak{so}(2,4)\oplus \mathfrak{u}(1)_R$.
Among its  maximal subalgebras (see 
   table 2 of \cite{Agmon:2020pde})  there is a unique   2-embedding\,\footnote{\,The other maximal subalgebra in the list  is $ \mathfrak{osp}(1\vert 4; \mathbb{R})$. Its
   bosonic component is  $ \mathfrak{so}(2,3)$, so it corresponds   to $p=3$  superconformal defects, i.e. interfaces or boundaries.
   }\vskip -5mm
     \bea
     \hskip -6mm \mathfrak{su}(2,2\vert 1) \supset 
   \mathfrak{su}(1,1\vert 1)\hskip -0.6mm\oplus \hskip -0.6mm  \mathfrak{su}(1,1)_{\rm c} \hskip -0.6mm\oplus \hskip -0.6mm \mathfrak{u}(1)_{\rm c}   \overset{bos}{\supset}  
   \mathfrak{so}(2,1 )\hskip -0.6mm\oplus \hskip -0.6mm \mathfrak{so}(2)_{\rm def}
    \hskip -0.6mm\oplus \hskip -0.6mm \mathfrak{so}(2,1)_{\rm c} \hskip -0.6mm\oplus \hskip -0.6mm \mathfrak{u}(1)_{\rm c}\, .  
   \eea
  Since a  chiral half  of the conformal algebra  commutes with the surviving  supercharges, these
  defects  have
   (2,0)  worldsheet supersymmetry. As was the case  for  line defects in three dimensions, the preserved   $\mathfrak{so}(2)_{\rm def}$ is   in general non-trivially embedded in  the bulk
   $\mathfrak{so}(2) \oplus \mathfrak{u}(1)_{R}$, while
      the residual  $\mathfrak{u}(1)_{\rm c}$ may    be broken. 
Surface  defects of this type  have been discussed in  refs.\,\cite{Koh:2009cj,Drukker:2017dgn,Razamat:2018zel}, and  the conjecture \eqref{conj} was proved for this case  in ref.\,\cite{Bianchi:2019sxz}.
  
 Extended  ${\cal N}> 1$ supersymmetry allows several different  maximal 2-embeddings,  
  \bea
    \mathfrak{su}(2, 2\vert {\cal N}) \ \supset\  \mathfrak{su}(1, 1\vert {\cal N}_+) \oplus \mathfrak{su}(1, 1\vert {\cal N}_-)\oplus \mathfrak{u}(1)_{\rm c}
\eea
 where ${\cal N}_++{\cal N}_-={\cal N}$ and $\mathfrak{su}(1, 1\vert 0)$\,=\,$ \mathfrak{su}(1, 1)$.\footnote{\,In the special case
   $({\cal N}_+, {\cal N}_-)$\,=\,$(2,2)$    the maximal 2-embedding is actually centrally extended as follows: 
$\,\,   \mathfrak{psu}(2, 2\vert 4)  \supset  \bigl[\mathfrak{u}(1) \rtimes \mathfrak{psu}(1, 1\vert 2) \rtimes \mathfrak{u}(1) \bigr]
    \oplus  \bigl[\mathfrak{u}(1) \rtimes\mathfrak{psu}(1, 1\vert 2) \rtimes \mathfrak{u}(1) \bigr]^\prime\ . 
$ 
}\,\,The   corresponding  superconformal  surface defects have  $2{\cal N}_+$  left-moving   and $2{\cal N}_-$  right-moving supercharges. 
All these   share   a  common  $\mathfrak{su}(1, 1\vert 1)$ subalgebra which is  sufficient for the proof of   the conjecture \eqref{conj}. 
 There is a large  literature treating various aspects of such defects, for a partial list of references see
 \cite{Gukov:2006jk,Gomis:2007fi,Gukov:2008sn,Gaiotto:2012xa,Gomis:2014eya,LeFloch:2015rpq}. 

   \vskip 1pt

Consider next the  six-dimensional SCFTs whose symmetry is    $\mathfrak{osp}(8^*\vert 2{\cal N})$,  where ${\cal N}=1$ or $2$. 
 The relevant maximal real  subalgebras are\,\footnote{\,When one of the even-integer  entries  in  $\mathfrak{osp}(2m^* \vert 2k)$
is  zero the algebra is purely bosonic,  
 $\mathfrak{osp}(0\vert 2k)\simeq \mathfrak{usp}(2k)$ and   $\mathfrak{osp} (2k^* \vert 0) = \mathfrak{so} (2k^*)$.
  The star in $\mathfrak{so} (2k^*)$
   indicates  the quaternionic real form of the complex Lie algebra $\mathfrak{so}(2k, \mathbb{C})$.  
 For small $k$ one has the following equivalences:
  $
 \mathfrak{so}(2^*)\hskip -0.5mm  \simeq \mathfrak{so}(2)\,; \hskip 0.5mm 
  \mathfrak{so}(4^*) \hskip -0.5mm \simeq \mathfrak{so}(2,1)\oplus \mathfrak{so}(3) \,; \hskip 0.5mm $
 $ \mathfrak{so}(6^*)\hskip -0.5mm \simeq \mathfrak{su}(1, 3)$ and $
\mathfrak{so}(8^*)\hskip -0.5mm \simeq \mathfrak{so}(2,6)\,. 
 $ Note also in passing the equivalences $\mathfrak{osp}(2^* \vert 2)\simeq \mathfrak{su}(2\vert 1)$  and 
 $\mathfrak{osp}(4^* \vert 2)\simeq D(2, 1; -2 ; 0)$  \cite{DHoker:2008wvd}. 
 }
\bea\label{6-2}
\mathfrak{osp}\bigl((8\hskip -0.4mm -\hskip -0.4mm 2m)^*\vert 2{\cal N}_1 \bigr) \oplus \mathfrak{osp}(2m^*\vert 2{\cal N}_2)\quad 
{\rm with}\quad {\cal N}_1+ {\cal N}_2 = {\cal N}, \ \  0\leq m\leq 4\ . 
\eea
For ${\cal N}=1$ there  is a unique  maximal 2-embedding 
\bea
\hskip -4mm 
\mathfrak{osp}(4^*\vert 2) \oplus \mathfrak{osp}(4^*)_{\rm c}
\ \overset{bos}{\supset} \ 
  \mathfrak{so}(2,1)\oplus \mathfrak{su}(2) \oplus \mathfrak{usp}(2) \oplus
   \mathfrak{so}(2,1)_{\rm c} \oplus \mathfrak{su}(2)_{\rm c}\ . 
\eea
A  chiral half  of   $ \mathfrak{so}(2,2) \simeq \mathfrak{so}(2,1)_+\oplus \mathfrak{so}(2,1)_-$  
commutes in this embedding  with the supercharges, so the
  half-BPS surface defects  have  (4,0)  worldsheet  supersymmetry.  
The defect R-symmetry is $\mathfrak{so}(4)_r\simeq \mathfrak{su}(2) \oplus \mathfrak{usp}(2)$, while the (unbroken)  rotation group is  
$\mathfrak{so}(4) \simeq \mathfrak{su}(2) \oplus \mathfrak{su}(2)_c$. 
  
 The  ${\cal N}=2$ superalgebra offers more possibilities  with  (8,0), (4,0) or (4,4) worldsheet supersymmetry\,. 
Examples of   (4,4) and (4,0)  defects are  self-dual strings  or  M2-branes ending on M5-branes  
  (for a partial list of  references see
 \cite{Howe:1997ue,Henningson:1999xi,Bachas:2013vza,Drukker:2020dcz,Faedo:2020nol,Conti:2024qgx}). 
%These defects   share  an  $\mathfrak{osp}(4^*\vert 2)$ superalgebra which suffices   for proving the conjecture.
The  (8,0) embedding,  on the other hand,  
is not   realized by non-trivial surface defects, i.e.  defects that  interact with  the bulk SCFT.  The reason is that  the putative  displacement operator
 should transform in  the  $(j, \, j^\prime)= ({1\over 2} ,\, {1\over 2} )$ representation
of $\mathfrak{so}(4)_{\rm def}\simeq \mathfrak{su}(2) \oplus \mathfrak{su}(2)^\prime$ and have  scaling dimension $\Delta = 3$\,.  
None of the unitary representations of 
  $\mathfrak{osp}(4^*\vert 4) $ (see table 8 of ref.\,\cite{Agmon:2020pde}) 
 includes  such an operator as its  
  top component. 
 
What is in any case  important for us here is that all   superconformal surface  defects in six dimensions  share  an  $\mathfrak{osp}(4^*\vert 2)\simeq 
\scalebox{0.94}{$D(2,1; -2;0)$} $ superalgebra which 
is sufficient for  proving  the conjecture \eqref{conj}.
 
 %%%%%%%%%%%%%%%%%   
   The last  theory  to consider is the ${\cal N}=1$  SCFT in five dimensions.  It 
admits the   maximal  2-embedding   
\bea\label{twoin5}
\hskip -4mm 
F(4;2) \, \supset\   D(2,1;2;0) \oplus   \mathfrak{so}(2,1)_c  \ \overset{bos}{\supset} \ 
  \mathfrak{so}(2,1)\oplus   \mathfrak{so}(4)    \oplus   \mathfrak{so}(2,1)_c\  ,  
\eea
which describes  surface defects with $(4,0)$ worldsheet supesymmetry. 
There seems to be   some  confusion   in the literature concerning   real subalgebras  of $F(4;2)$,\footnote{
 In particular, the  first $F(4;2)$ entry in table 2 of 
  \cite{Agmon:2020pde}  
  cannot be correct, since  the fermionic generators in  $ \mathfrak{su}(2\vert 1)\hskip -0.4mm  \oplus \hskip -0.4mm  \mathfrak{su}(1, 2)_{\rm c} $  
    would commute with the dilatation operator. 
   }
   so  we  work out  this case in
 detail in appendix \ref{F42}\,.
Holographic surface defects of this type were  found in  gauged suspergravity in ref.\,\cite{Karndumri:2024gtv}.

To summarize, half-BPS  superconformal and rotation-invariant surface defects exist for ${\cal N}=1$ theories in $n=4,5$ and 6 dimensions. 
The preserved  superalgebras,  modulo bosonic commutants,   
are $\mathfrak{su}(1, 1\vert 1)$, $ D(2,1;2;0)$
and $\mathfrak{osp}(4^*\vert 2)$\,.

  %%%%%%%%%%%%%%%%%%%%%%%%%%%%%%%
 %%%%%%%%%%%%%%%%%%
 \subsection{Higher-${p}$  defects} 
 
    Finally we  consider  $p=3, 4$.  Stand-alone, superconformal  $p=3$  defects do not 
     exist in $n=6$  dimensions because none of the maximal subalgebras of $ \mathfrak{osp}(8^*\vert 2{\cal N})$
is  an odd-$p$ embedding \cite{Agmon:2020pde}. 
A maximal 3-embedding does exist in  $n=5$ dimensions
  \bea
\hskip -4mm 
F(4;2) \, \supset\  \mathfrak{osp}(2\vert 4; \mathbb{R}) \oplus  \mathfrak{u}(1)_{\rm c}  \  \overset{bos}{\supset}\   \mathfrak{so}(2,3) \oplus \mathfrak{so}(2)_{\rm def} \oplus 
 \mathfrak{u}(1)_{\rm c} \ . 
 \eea
 As with other  codimension-two defects, the preserved transverse rotations  are non-trivially embedded in the bulk 
 $\mathfrak{so}(2)  \oplus \mathfrak{u}(1)_{R}$. 
Holographic defects of this type have been analyzed in ref.\,\cite{Gutperle:2020rty,Santilli:2023fuh}.\footnote{\, Despite the title  
of \cite{Gutperle:2020rty}, the defects  analyzed  in this paper are
 three-dimensional.}

     The last case is $p=4$, $n=6$  corresponding to the   maximal 4-embedding  
 \bea
  \mathfrak{osp}(8^*\vert 2{\cal N}) \,\supset\,  \mathfrak{su}(2,2\vert {\cal N}) \oplus \mathfrak{u}(1)_{\rm c} \,\overset{bos}{\supset}\,  \mathfrak{so}(2,4) 
  \oplus \mathfrak{so}(2)_{\rm def}\oplus \mathfrak{su}({\cal N})\oplus \mathfrak{u}(1)_{\rm c} \,. 
  \eea
 These  defects are half-BPS, with the  transverse rotations once again non-trivially embedded in the  bulk R-symmetry. 
The gravity duals of such holographic defects for ${\cal N}=2$   are expected to  belong to the general class of M-theory solutions found  in 
 refs.\,\cite{Maldacena:2000mw,Gaiotto:2009gz}. 
The solutions were  mainly  considered as  AdS$_5$ compactifications,  i.e. as autonomous four-dimensional SCFTs, and
  it would be very  interesting to  explore  them  further as  $n=6$ DCFTs along the lines of refs.\,\cite{Ferrero:2021wvk, Gutperle:2022pgw, Gutperle:2023yrd}.
  % correct2: references added
 
% Note   that the unbroken superalgebras  all admit a conserved stress-tensor multiplet. This is consistent with the fact that  decoupled $p$-dimensional SCFTs can  always be added at no cost
% on the worldvolume of the defect. 
 
\vskip 1pt   
 This completes our derivation of table \ref{tab1}. The  last thing we need to check is that  it is enough to
   prove  \eqref{conj} for the entries of this table. One may have worried,   for example,   that 4d ${\cal N}=4$ super Yang-Mills has 1/4-BPS  defects
  that cannot be expressed as half-BPS defects of  a ${\cal N}=2$ theory. For line defects this possibility  is ruled out by inspection of  the complete list of ref.\,\cite{Agmon:2020pde}. 
 For surface defects in four  and six dimensions   there exist no  non-trivial 2-embeddings in  the extended superconformal algebras ${\cal G}_s^{(n)}$ that are not also embedded  in  a
 ${\cal N}=1$ subalgebra of 
${\cal G}_s^{(n)}$. 
  Finally, the only 4-embeddings  in the 6d  ${\cal N} = (2, 0)$ theory are $\mathfrak{su}(2,2\vert 2)$,  or its unique up to isomorphisms  subalgebra $\mathfrak{su}(2,2\vert 1)$.

 %%%%%%%%%%%%%%%%%%%%%%%%%%%%%%%%%%%%%%%%%%%%%%%%%%%
 %%%%%%%%%%%%%%%%%%%%%%%%%%%%%%%%%%%%%%%%%%%%%%%%%%%

 \section{An equivalent conjecture}\label{TDcor}
 
  The conjecture \eqref{conj} relates the one-point function of the stress tensor  $\langle\hskip -1mm \langle T^{\mu\nu} \rangle\hskip -1mm\rangle$ to the
  two-point function of the displacement   $\langle\hskip -1mm \langle D^i   D^j \rangle\hskip -1mm\rangle $\,. 
  In this section  we will  reexpress it as a condition on  the bulk-to-defect two-point function  $\langle\hskip -1mm \langle T^{\mu\nu} D^j \rangle\hskip -1mm\rangle$\,. 
 Here and in what follows\, 
$\langle\hskip -1mm \langle \cdot \cdot  \cdot \rangle\hskip -1mm\rangle$\, stands for   the normalized  correlation functions  in the background  of a static (hyper)planar defect.

%%%%%%%%%%%%%  
\subsection{The  displacement-stress tensor  correlation}\label{TDco}
  The two-point function of any bulk conformal primary   with the displacement   is strongly constrained by 
 conformal invariance  \cite{Billo:2016cpy}. Consider as a warmup a scalar operator ${\cal O}$ with scaling dimension $\Delta_{\cal O}$\,,
 and with one-point function 
 \bea\label{O1pt}
 \langle\hskip -1mm \langle {\cal O}(x) \rangle\hskip -1mm\rangle \, = \, {a_{\cal O}\over \vert x_\perp\vert^{\Delta_{\cal O}}}\ . 
 \eea
 Its  two-point function with the displacement is completely fixed by this data 
 \bea\label{OD}
\hskip -6mm \langle\hskip -1mm \langle   {\cal O} (x) D^j(y)\rangle\hskip -1mm\rangle
=  b_ {\cal O} \, 
{ x^j \vert x_\perp\vert^{p- \Delta_{\cal O}}\over \vert x\min y \vert^{2(p+1)}}\ , \quad
 %\vskip -7mm
 {\rm with}\quad  b_{\cal O} =  \, 2^p\, \Gamma ({p+1\over 2} )  \pi^{-(p+1)/2} \,  \Delta_{\cal O} a_{\cal O} \  . 
\eea 
Our notation is as follows: $x= (x^\alpha, x^j)$  and $y=(y^\alpha, 0)$ 
 where early Greek letters  
label    directions along the defect,  
$\alpha  \in \{0, \cdots , p$$-$$1$\},  
while   middle  Latin letters  label  the directions transverse to the defect,  
$j\in\{ p, \cdots , n$$-$$1\}$\,. 
Furthermore
   $\vert x\vert^2 =  \eta_{\mu\nu}x^\mu x^\nu$ and 
 $ \vert x_\perp\vert^2 = \delta_{ij} x^i x^j$. 
 
  The general form \eqref{OD} is fixed by the unbroken $\mathfrak{so}(2,p)$\,$\oplus$\,$\mathfrak{so}(n-p)$ invariance. The coefficient $b_ {\cal O}$  follows from the identity
 \bea\label{eucl} 
 \int d^p y  \,  \langle\hskip -1mm \langle   {\cal O} (x) D^j(y)\rangle\hskip -1mm\rangle \,=\, 
 - {\partial \over \partial x^j}  \langle\hskip -1mm \langle {\cal O}(x) \rangle\hskip -1mm\rangle \, 
 \eea
 which expresses the fact that the operator $\exp(i a\int d^p y\, D^j (y))$ translates  the defect   in the $j$th 
 direction by an amount $a$.\footnote{\,The integral in  \eqref{eucl}
 is well-defined  in the Euclidean theory.
 }
 This identity is valid for any operator,  scalar or  tensor,    descendant or primary.

Consider next a symmetric traceless tensor $T^{\mu\nu}$. 
The unbroken conformal and transverse-rotation symmetry,  $\mathfrak{so}(2,p)\oplus  \mathfrak{so}(n$$-$$p)$,   
determines  $\langle\hskip -1mm \langle T^{\mu\nu} D^j \rangle\hskip -1mm\rangle$ in terms of three parameters
  $\mathfrak{b}_{1,2,3} $\,. 
 Explicitly 
     \cite{Billo:2016cpy, Bachas:2024nvh}
     \bea\label{physco}  
\langle\hskip -1mm \langle T^{\mu\nu}(x)  D^j (y)\rangle\hskip -1mm\rangle
 \,= \,  {
  \vert x_\perp\vert^{p-\Delta_T}  \over \vert x\min y\vert^{2p+2}} 
\,  G^{\mu\nu;\, j}(x,y)\  \quad {\rm with}
\eea
%%%%
\vskip -5.5mm 
 \begin{subequations}\label{physcorr}
\begin{align}
&G^{\,\alpha\beta;\, j}  =\,
 {1\over n} \bigl( (n\min p\min 1)\mathfrak{b}_2 - \mathfrak{b}_1 
  \bigr)\, { \eta^{\alpha\beta}  x^j } + \, 4\mathfrak{b}_1 \, { (x\min y)^{\alpha}(x\min y)^{\beta}x^j \vert x_\perp\vert^2  \over  \vert x\min y\vert^4  }\,
 \,\, ;   \label{physcorr1}
\\
&G^{\, i \beta ;\, j} = 
 {
\,   
}   -\mathfrak{b}_3 \, { \delta^{ij}  {\vert x_\perp\vert^2 (x\min y)^\beta \over \vert x\min y\vert^2}   } 
 +  (   \mathfrak{b}_3 \min 2 \mathfrak{b}_1 
  )\,
 {   x^i x^j (x\min y)^\beta \over   \vert x\min y\vert^2 }
+  \scalebox{0.92}{$ 4\mathfrak{b}_1$} \,
 {   x^i x^j (x\min y)^\beta\vert x_\perp\vert^2  \over   \vert x\min y\vert^4 }\,
 \,\, ; \label{physcorr2}
\\
&G^{\,ik;\, j}  =\,
 -{1\over n} \bigl(   (p+1)\mathfrak{b}_2 + \mathfrak{b}_1 
  \bigr)\,
\delta^{ik}  {x^j } +  
 {\mathfrak{b}_3\over 2}    \, (\delta^{ji } x^{k}+ \delta^{jk } x^{i}
 ) \bigl(
  {\scalebox{0.99}{$1$}     } -  
  { \scalebox{0.87}{$2$} \,\vert x_\perp\vert^2 \over  \vert x\min y\vert^2    }\bigr)
  \nonumber \\
 &\, \hskip 0.8cm  + ( \mathfrak{b}_1 + \mathfrak{b}_2 - \mathfrak{b}_3)
   { x^i x^k x^j  \over    \vert x_\perp\vert^2} 
 +  ( 2\mathfrak{b}_3 -4 \mathfrak{b}_1 
  )\,  { x^i x^k x^j  \over     \vert x\min y\vert^2}
 + 4\mathfrak{b}_1  \, { x^i x^k x^j  \vert x_\perp\vert^2 \over    \vert x\min y\vert^4} \,\,
 .  \label{physcorr3}
 % correct2: removed extra (x) in 3.5c
  \end{align} 
\end{subequations}
 \smallskip
The identity \eqref{eucl} with ${\cal O}$ replaced by $T^{\mu\nu}$ fixes two of the parameters, 
  \bea\label{wardd1}
(p+1) \mathfrak{b}_2 + \mathfrak{b}_1\, = \,    {\Delta_T \over 2} \mathfrak{b}_3    \  \quad {\rm and} \quad 
\mathfrak{b}_3 = 2^{\,p+2} \pi^{-(p+1)/2}\Gamma ( { \scalebox{0.92}{$p+3$} \over 2} )\,a_{\rm T}\ ,  
\eea
so  one  parameter 
  stays free. But in 
  the special case where   $T^{\mu\nu}$ is the SCFT  energy-momentum tensor, $\Delta_T =n$
 and    the broken conservation 
 law $\partial_\mu T^{\mu j} = \delta(x_\perp) D^j$ 
% \eqref{conserve} 
gives one more  equation that relates   the remaining parameter
to  the displacement norm,\,\cite{Billo:2016cpy}  
\bea\label{wardd2}
2p\,\mathfrak{b}_2 - \scalebox{0.92}{$ (2n-p-2) $} \, \mathfrak{b}_3\ =  \ { (n-p) \Gamma({n-p\over 2})\over
\pi^{(n-p)/2 }}\, C_D
 \ . 
\eea
Taken together  the relations \eqref{wardd1} and \eqref{wardd2} can now be used 
to determine $\langle\hskip -1mm \langle T^{\mu\nu} D^j \rangle\hskip -1mm\rangle$ completely   in terms of the  DCFT data $a_{\rm T}$ and $C_D$.
%\smallskip

Since the  conjecture  \eqref{conj} relates $C_D$ to $a_T$,  we can  re-express it  as a relation between the parameters 
$\mathfrak{b}_{1,2,3}$. It  takes the 
 elegant  form
  \smallskip  
\bea\label{conjj}
 \boxed{2n\mathfrak{b}_1 = (p+2)\mathfrak{b}_3\ .} 
\eea  
%\vskip 3pt \noindent    Alternatively, one could   trade  $a_T$ and $C_D$ for two new parameters,  $b_T$ and $\delta$,  in terms of which the solution of eqs.\,\eqref{wardd1}\,-\,\eqref{wardd2}
%can be written as
%\bea
%\mathfrak{b}_1 = (p+2) b_T+ n \delta \ , \quad  (p+1) \mathfrak{b}_2 = \bigl({n^2 \hskip -0.9mm -\hskip -0.6mm p\hskip -0.6mm -\hskip -0.6mm 2}\bigr)\, b_T   \ ,
% \quad \mathfrak{b}_3 = 2n\,b_T +  2\delta \  .  
%\eea
We have thus succeeded in reformulating the original conjecture as a condition on the bulk-to-defect correlation function 
$\langle\hskip -1mm \langle T^{\mu\nu} D^j \rangle\hskip -1mm\rangle$ which, as we will prove in section \ref{sec:proof}, follows from superconformal Ward identities.

%%%%%%%%%%%%%%%%%%%%%%%%%%%%%%%%%%%%%%%%%%%%%%%%%%%%

  \subsection{Line defects in ${\cal N}=2$ abelian gauge theory}\label{sec:abelian}
  To put some flesh into these  abstract formulae, and to prepare the ground for a discussion of the underlying physics, 
 let us  consider  the  example of superconformal  line defects 
in  the  ${\cal N}=2$ abelian gauge theory in four dimensions.

  The ${\cal N}=2$  vector multiplet  contains  the gauge field $A_\mu$,  two Weyl photinos  and two real scalar fields $\phi_{I=1,2}$\,.  
 The  general (electrically-charged) conformal  line defect is  described by  the operator
 \bea\label{W}
 W = \exp  \bigl( \int \hskip -0.4mm ds\, (ie  A_\mu \dot y^\mu + g \, \vert \dot y\vert \hat n^I \phi_I ) \bigr) \ , 
 \eea
 where $\hat n^I\hat n_I=1$, dots are $s$-derivatives  and the bulk fields are evaluated at $y^\mu(s)$.  The unit vector 
  $n^I$ could in principle depend on   $s$,  but we here take it  constant and write for short $\hat n^I \phi_I = \phi$. 
\vskip 1pt

 The stress tensor  of the bosonic fields is $T_{\mu\nu} = T_{\mu\nu}^{(s)} + T_{\mu\nu}^{(v)}$ where
      \begin{subequations}\label{B26}
\begin{align} 
% correct: lowered indices in (3.10a)
 &T_{\mu\nu}^{(s)} = \partial_\mu\phi \partial_\nu\phi - {1\over 2} \eta_{\mu\nu}\vert \partial\phi\vert^2 + {1\over 6}\,  (\eta_{\mu\nu} \square - \partial_\mu \partial_\nu )\phi^2 \ , 
\label{Tscl}
 \\
 &
  \hskip 1cm  {\rm and} \qquad T_{\mu\nu}^{(v)} = F_\mu^{\ \rho} F_{ \nu\rho}  - {1\over 4} \eta_{\mu\nu} \vert F\vert^2  \ .     
 \end{align} 
\end{subequations}     
 The scalar stress tensor  $T_{\mu\nu}^{(s)}$ includes the total-derivative  term that makes it  traceless.
To find the displacement operator we parametrise  the worldline in  static gauge,  $(y^0, \ y^j ) = (s, y^j (s))$\,, and
expand \eqref{W} to linear order in $y^j$.  The result is 
 \bea
  D^j = e F^{0j}  +  g\,  \partial^j  \phi\ . 
 \eea
 Finally,  the classical background fields created by  the static defect read
 \bea\label{B4}
\phi_{\rm class}  =   {g\over 4\pi\vert x_\perp\vert}\ , \quad {\rm and} \quad
 F^{0j}_{\rm class}   =  {ex^j\over 4\pi\vert x_\perp\vert^3} \ . 
 \eea
 Since the  DCFT at hand  is free, eqs.\,\eqref{B26}\,-\,\eqref{B4} is all  that we need   to  calculate any  correlation function.

The one-point function $ \langle\hskip -1mm \langle \, T^{\mu\nu}  \rangle\hskip -1mm\rangle$ is given entirely by  the  classical fields. Comparing to the general form  eq.\,\eqref{26} we find
 \bea
 {a^{(s)}_T} =  - {g^2\over 48\pi^2} \ \ ; 
 \qquad 
 {a^{(v)}_T } =  - {e^2\over 16\pi^2} \ . 
 \eea
From  the scalar propagator $\Delta(x-y) = 1/4\pi^2 \vert x-y\vert^2$   we furthermore get
 \bea
 {C_D^{(s)}} =   {g^2\over 2\pi^2} \ \ ; 
 \qquad 
  {C_D^{(v)}}  =  {e^2\over \pi^2} \ . 
 \eea
For  $p=1, n=4$ 
the conjecture \eqref{conj} 
   reads $C_D = -18 a_T$. This is not valid separately for  the photon or the scalar. But
 the half-BPS line defect  has $e=g$,\footnote{\,There also exist $1/4$-BPS  defects but these  break the transverse-rotation symmetry  \cite{Zarembo:2002an}.}
  and   one can check that the conjecture  is indeed satisfied.

 To calculate $ \langle\hskip -1mm \langle \, T^{\mu\nu} D^j  \rangle\hskip -1mm\rangle$ 
 we contract the displacement with one of the two free fields in the stress tensor, and replace 
 the remaining  field by  its classical value. 
  In the Lorentzian theory   $\Delta(x-y)$  is the Feynman propagator if  $y^j(s)$ is treated as a quantum field, or the retarded one
if it is treated as  a classical background.
 Since the general form \eqref{physco}\,-\,\eqref{physcorr} relied only on conformal symmetry,   the result  has  this  form separately for the scalar  and the vector. One finds
      \begin{subequations}\label{Bbs}
\begin{align}    
 & \mathfrak{b}_1^{(s)} = -{g^2 \over 4\pi^3} \ ,  \quad \mathfrak{b}_2^{(s)} = -{g^2 \over 24\pi^3}  \ ,  \quad \mathfrak{b}_3^{(s)} = -{g^2 \over 6 \pi^3}\ ;  
 \\\label{Bbs2}
  &\mathfrak{b}_1^{(v)} = 0\ ,  \quad \mathfrak{b}_2^{(v)} = -{e^2 \over 2\pi^3} \ ,  \quad \mathfrak{b}_3^{(v)} = -{e^2 \over 2\pi^3} \ . 
 \end{align} 
\end{subequations}    
The conjecture \eqref{conjj},  which is equivalent to \eqref{conj},  reads $8 \mathfrak{b}_1= 3 \mathfrak{b}_3$. 
%$8 (\mathfrak{b}_1^{(s)}+\mathfrak{b}_1^{(v)})= 3 (\mathfrak{b}_3^{(s)}+\mathfrak{b}_3^{(v)})$. 
Not surprisingly, it is not valid for the scalar and the vector separately, but  if  $e=g$ it is indeed  satisfied,
$8 (\mathfrak{b}_1^{(s)}+\mathfrak{b}_1^{(v)})= 3 (\mathfrak{b}_3^{(s)}+\mathfrak{b}_3^{(v)})$\,.

 %%%%%%%%%%%%%%%%%%%%%%%
%%%%%%%%%%%%%%%%%%%%%%%%%%%%%%%%%%%%%%%%%%%%%%%%%%%%%%%%%%%%%%%%%%%%%%%%%%%%%%%%%%%%
%%%%%%%%%%%%%%%%%%%%%%%%%%%%%%%%%%%%%%%%%%%%%%%%%%%%%%%%%%%%%%%%%
  \section{Stress, radiation and the Schott term}\label{sec:4}
    
    Despite its simplicity, the new form \eqref{conjj} of the  conjecture cannot yet  be exploited for a proof.  
 Furthermore, its physical meaning is obscure. In search of  inspiration let us return  to  the original argument of Lewkowycz and Maldacena 
  \cite{Lewkowycz:2013laa} that motivated the relation between $C_D$ and $a_T$.  

 These  authors   wanted 
    to reconcile  two different  calculations\footnote{\,For  earlier
     calculations of radiation from the holographic-dual string in AdS see  refs.\,\cite{Mikhailov:2003er,Athanasiou:2010pv,Hubeny:2010bq,Hatta:2011gh,Baier:2011dh}. 
      } 
       of the energy emitted  by an accelerating   
      half-BPS `quark'   in the  ${\cal N}=4$ super Yang-Mills theory.
      One calculation used  the Bremstrahlung function which is proportional to $C_D$ \cite{Correa:2012at}, while  the other   
      computes the energy flux  directly  in the background of a uniformly-accelerating quark  which is conformal  to a static one \cite{Fiol:2012sg}.
       The discrepancy between the  two results was attributed to the difficulty of
      separating  the  radiation  from the  self-energy of the defect.  
This is a time-honoured  problem  even  in  classical electrodynamics \cite{Dirac,FR,LL,Teitelboim:1970mw,Poisson:1999tv,Galtsov:2004uqu}.

     Noting that the ${\cal N}=4$ theory has a scalar operator ${\cal O}$ with scaling dimension $\Delta_{\cal O} =n-2$ which couples  to the defect, Lewkowycz and Maldacena
      proposed that the emitted radiation should be   computed by the  modified  stress tensor  
     \bea\label{modstress}
\tilde T^{\mu\nu}  = T^{\mu\nu}\,  +  \xi \,  (\eta^{\mu\nu} \square - \partial^\mu \partial^\nu ) {\cal O}\   
\eea
for an appropriate value of $\xi$.  The new stress tensor  is conserved but not  traceless. 
 The intuition in ref.\,\cite{Lewkowycz:2013laa}, further elaborated in
 \cite{Agon:2014rda,Fiol:2015spa,Fiol:2019woe,Bianchi:2019dlw},
  was that choosing $\xi$  so as to remove the leading short-distance singularity 
 of  $\tilde T^{\mu\nu}$ near  the defect would  also neatly separate  the defect's self-energy. 

 The  intuition is indeed  correct,  but the role of supersymmetry  has remained unclear.
In this section we propose  a more detailed  and physical explanation.

%%%%%%%%%%%%%%%%%%%%%%%%%

\subsection{Static stress and the NEC}\label{sec:4.1}
 
      The leading singularity 
 of  $\tilde T^{\mu\nu}$  near the defect is given by its one-point function. Using the general form \eqref{O1pt} for  $ \langle\hskip -1mm \langle {\cal O}(x) \rangle\hskip -1mm\rangle$  
 we find
    \begin{subequations}\label{T1pt}
\begin{align}    
 &
\langle\hskip -1mm \langle\,  \tilde T^{\alpha\beta}  \rangle\hskip -1mm\rangle = \langle\hskip -1mm \langle\,  T^{\alpha\beta}  \rangle\hskip -1mm\rangle
\,+\, \xi a_{\cal O} \, p   (n\hskip -0.5mm - \hskip -0.5mm2)\,   
{\eta^{\,\alpha\beta}\over \vert x_\perp\vert^n}
\ , \qquad
\langle\hskip -1mm \langle \, \tilde T^{\,i \beta }   \rangle\hskip -1mm\rangle = \langle\hskip -1mm \langle\,  T^{\,i \beta }   \rangle\hskip -1mm\rangle = 0\  
\\ 
&
{\rm and} \qquad \langle\hskip -1mm \langle\,  \tilde T^{ik}  \rangle\hskip -1mm\rangle \,= \, \langle\hskip -1mm \langle\,  T^{ik}  \rangle\hskip -1mm\rangle\,
+ \, \xi a_{\cal O}  
(n\hskip -0.5mm - \hskip -0.5mm2) 
\biggl[   (p+ \hskip -0.5mm 1){\delta^{ik}\over \vert x_\perp\vert^n} -  n\, {x^i x^k\,\over \vert x_\perp\vert^{n+2}} \,  
 \biggr] \ .  
\end{align} 
 \end{subequations}
Comparing to  expression \eqref{26} for 
  $ \langle\hskip -1mm \langle\,    T^{\mu\nu}  \rangle\hskip -1mm\rangle$ shows  that it is impossible to remove the singularity in both the worldvolume  and  transverse components.
Since the energy better stay positive,  we  should  try  to remove the latter singularity. 
This is indeed possible with the choice $\xi = \xi_{\rm def}$, where  
\bea\label{xi}
  \xi_{\rm def}\, a_{\cal O} \, 
 =\,  {a_T \over n (n\hskip -0.5mm - \hskip -0.5mm2)}\ . 
\eea
The  subscript `def'   is here to remind us that the above  choice of $\xi$ depends on the defect via the DCFT data $a_{\cal O}$ and $a_T$.

Inserting \eqref{xi} in \eqref{T1pt} one  finds
%then $\langle\hskip -1mm \langle  \tilde T_{i j }  \rangle\hskip -1mm\rangle =0$.  
    \bea\label{312}
 \langle\hskip -1mm \langle\,   \tilde T^{\,\alpha\beta}  \rangle\hskip -1mm\rangle
\,=\, 
 a_{T}   \bigl( 1-
{1\over n}\bigr)\, 
{\eta^{\,\alpha\beta}\over \vert x_\perp\vert^n}\ \ , 
%\qquad {\rm and}  
\qquad
\langle\hskip -1mm \langle\,  \tilde  T^{\,i \beta }  \rangle\hskip -1mm\rangle\, = \, 
\langle\hskip -1mm \langle\,  \tilde  T^{\,i k}  \rangle\hskip -1mm\rangle\, = \, 0\ . 
  \eea
The key point here  is that,  with the above choice of $\xi$, 
%because  the fields of  a static (hyper)planar defect are free of   stress, 
 the singularity  of $\tilde  T^{\mu\nu }$ at the defect can be absorbed by a  redefinition of  the defect's tension or mass. For  a line defect with proper-time $\tau$   for example 
\bea
 T^{\mu\nu }_{\rm tot}(x) \, = \,  m\hskip -0.4mm  \int d\tau\, \delta^{(n)}(x\min  y)\, {dy^\mu\over d\tau} {dy^\nu \over d\tau} \,+\,  \tilde  T^{\mu\nu }_{\rm reg}(x)
\eea
where $y^\mu(\tau)$ is the worldline of the defect, $m$ its renormalised mass and $\tilde T^{\mu\nu }_{\rm reg}$ has no $\vert x-y\vert^{-n}$ singularity. 
This  would not have worked  with $T^{\mu\nu }$ because  the singular  stress  $\langle\hskip -1mm \langle\,   T^{ik}  \rangle\hskip -1mm\rangle $ cannot be absorbed in $m$.
\vskip 1pt

   Two questions arise immediately. Who has ordered the modification of the bulk stress tensor  \eqref{modstress}, and  why do we  need  supersymmetry\,?
The  free-field example  of the previous section provides the answers. 
In this example the  scalar operator  ${\cal O}$ 
 is $\phi^2$\, and  its  one-point function is $\langle\hskip -1mm \langle\,  \phi^2   \rangle\hskip -1mm\rangle\, = \, {g^2/ 16\pi^2 \vert x_\perp\vert^2}$\,,
 so to cancel   the transverse stress we must choose
\bea
 \xi_{\rm def} =  {a_T\over 8 a_{\cal O}} =  - {g^2+3 e^2\over 24 g^2}\ .  
\eea
This can be done for arbitrary $e$ and $g$. But something special happens when 
 $e=g$, i.e. for supersymmetric  defects.  
  In this case  $\xi = -1/6$ which is  precisely the value needed to  remove the  total-derivative term in   
 \eqref{Tscl}. This term comes from the conformal $R\phi^2$  coupling of the  scalar field  
  and  is known  to violate  even the weakest of local energy conditions, 
the Null Energy Condition (NEC), see e.g.\,\cite{Visser:1999de}. It was indeed  shown in 
 ref.\,\cite{Fiol:2019woe}  that for $\xi\not= -1/6$ an accelerating defect coupling to $\phi$   emits NEC-violating radiation.\footnote{  But  the A(verage) NEC is  satisfied if 
 the defect  stops accelerating at
 $\tau \to\pm\infty$\,.}
 
 \vskip 1pt

The above   free-field example motivates  the following\,:
  \begin{center} %\vskip -5mm
\fbox{%
    \begin{tabular}{@{}c@{}}
  The  subtraction that restores  the NEC for  the bulk  energy-momentum  tensor\\
 also  removes the transverse stress of  static planar superconformal  defects.
       \end{tabular}%
    }
\end{center}
\vskip 3pt

We  will now explain why  this  assertion  is  generally valid. 
 %To this end we must first understand how  to  restore the NEC   in  a generic  DCFT.

%%%%%%%%%%%%%%%%%%%%%%%%%%%%%%%%%%%%%%%%%%%%%
%%%%%%%%%%%%%%%%%%%%%%%%%%%%%%%%%%%%%%%%%%%%%

\subsection{Energy flux of  a moving defect}\label{sec:moving}
  Ever since Dirac's seminal paper  \cite{Dirac} the question of radiation reaction for a  moving charge has been the subject 
  of   controversy. 
  The problem can be summarized as follows:  The energy flux in  the   fields created by  the charge 
 is the sum of  a radiation and a `bound-field' part  \cite{Teitelboim:1970mw,Galtsov:2004uqu}. These  are separately conserved. 
 Integrating the radiation flux   in electrodynamics gives  the standard Larmor formula which   is proportional to the acceleration squared. 
 In extensions of Maxwell's theory, such as the ${\cal N}=2$ supersymmetric theory of section \ref{sec:abelian},  the charge may also couple to scalar fields. 
 In this case  there can be a NEC-violating radiation  flux   proportional
 to  the derivative of the   acceleration  \cite{Fiol:2019woe}.
 
   The `bound-energy'  flux, on the other hand, is singular near the defect. The lore  is  that  this  singular flux  can  be absorbed 
  by  redefining    the particle   mass, and that the leftover  is  the 
   Schott  term in  the Lorentz-Dirac equation of motion. This term is also proportional to the derivative of the acceleration. 
  Its  necessity  in the Lorentz-Dirac equation follows from  a heuristic  argument of  Landau and Lifshitz \cite{LL}. 
     But calculating it   has remained  a challenge, see e.g. ref.\,\cite{Galtsov:2004uqu}.

    As anticipated above,   the two problems are related. 
    The operator that sets  the defect in motion is  $\exp ( i \hskip -0.5mm \int ds\, y^j(s) D^j(y))$ where  $s=y^0$  in static gauge. The  effects of interest   are linear 
    in the  acceleration, so they  can be  extracted  from
     $\int   \langle\hskip -1mm \langle   T^{i 0}  D^j   \rangle\hskip -1mm\rangle\,y^j $\,.
     The bulk-to-defect 
  2-point function   is given by  \eqref{physco}\,-\,\eqref{physcorr2}\,, we  copy it  here for the  reader's convenience:
 \bea\label{46E}
 % correct: from (x-y)^\beta to x^0 - y^0
  \langle\hskip -1mm \langle   T^{i 0}  D^j   \rangle\hskip -1mm\rangle = 
   {    (x^0\min y^0) \vert x_\perp\vert^{p -n}\over   \vert x\min y\vert^{2p+4} } 
   \Bigl(
   -\mathfrak{b}_3 \, { \delta^{ij}  {\vert x_\perp\vert^2 }   } 
 +  (   \mathfrak{b}_3 \min 2 \mathfrak{b}_1 
  )\,
 {   x^i x^j   }
+  \scalebox{0.92}{$ 4\mathfrak{b}_1$} \,
 {   x^i x^j  \vert x_\perp\vert^2  \over   \vert x\min y\vert^2 }
 \Bigr)\ . \ \ 
 \eea
 It has  singularities 
   when $x-y$ is null, or  when  $\vert x_\perp\vert$\,=\,$ 0$\,.

The NEC-violating  radiation  comes from  the leading $\vert x\min y \vert$   singularity which we can write as follows
   %which is indeed proportional to $ \mathfrak{b_1}$,  
  \bea\label{46l}
    \langle\hskip -1mm \langle   T^{i 0}(x) D^j(y)  \rangle\hskip -1mm\rangle\, =  \, 
 -{32 \, \mathfrak{b}_1\, \vert x_\perp\vert^{p+2-n} \over p(p\plus 1)(p\plus 2)}\, \partial^0 \partial^i \partial^j \, \bigl({\,1 \over \vert x\min y\vert^{2p}}\bigr) \ + \ {\rm subleading}. 
  \eea
  %%%%%%%%%%%%%
  
Note that this term is absent in Maxwell's theory, which obeys the NEC and for which  $\mathfrak{b}_1=0$. 
In the free-field example of  section \ref{sec:abelian} such a term can be seen to arise when we  contract  the displacement
with one of the two scalar fields in $T^{i0} \sim \partial^i \partial^0\phi^2$. 
The relevant Lorentzian Green function  is
 the retarded one
\bea\label{retG} 
{1\over 4\pi^2 \vert x-y\vert^{2} } 
  \ \rightarrow\ \Delta_R(x-y)\ = \ {i\theta(x^0\min y^0) \over 2\pi} \,\delta\bigl(\vert x\min y\vert^2\bigr)\ ,  
  \eea
  and all three derivatives act on it. 
Converting  $\partial^i \partial^j$  
  to  derivatives with respect to $s$   as in the familiar   calculation of the Li\'enard-Wiechert potential, 
and
integrating
%\,   $ \langle\hskip -1mm \langle   T^{i 0}  D^j   \rangle\hskip -1mm\rangle\,y^j(s) $
by parts gives the   NEC-violating  radiation found in \cite{Fiol:2019woe}.

%%%%%%%%%%%%%%%%  
  
Generalising  to an interacting theory  sounds  daunting, but   fortunately conformal 
     invariance  comes here to the rescue.  What we want is to  
    replace   the Euclidean  correlator \eqref{46E} by  the retarded 
    one, 
    \bea
       \langle\hskip -1mm \langle  T^{i 0}(x) D^j(y) \,\rangle\hskip -1mm\rangle\ \rightarrow \ 
      \langle\hskip -1mm \langle   \theta(x^0\min y^0)\,[\,T^{i 0}(x) , D^j(y)\,] \,\rangle\hskip -1mm\rangle\ .
      \eea
    In  a conformal theory  this can be  done    by a simple  
    % $i\epsilon$-prescription for the time coordinates
    analytic continuatio of Euclidean time.\footnote{\,For reviews of Lorentzian CFT see  \cite{Hartman:2015lfa,SDuff}\,.} 
    Explicitly,  the ordered $n$-point function\,  $\langle\hskip -1mm \langle  {\cal O}_1(x_1)\ \cdots \ {\cal O}_n(x_n)\rangle\hskip -1mm\rangle$\,  
  is given  by  the analytic  continuation  $x^0_r\to x^0_r- i\epsilon_r$ with $\epsilon_1>\epsilon_2 \cdots > \epsilon_n>0$. 
 For  line defects ($p=1$)  this prescription  leads precisely  to  the  replacement of \eqref{retG} in \eqref{46l}.
 This shows that, apart from  the  unknown   coefficient $\mathfrak{b}_1$, 
   the  NEC-violating radiation  is universal -- the same in all dimensions,
  and in  free as well as strongly-interacting DCFTs.\footnote{The reader may wonder why the  propagator of a free  scalar field $\phi$ in dimensions $n\not=4$ does not enter in 
  the correlator \eqref{46l}. The reason is that conformal defects do not  couple linearly to $\phi$ in other dimensions.
  }
  This agrees with the holographic calculation of `heavy quark' radiation in ${\cal N}=4$ super Yang-Mills at strong coupling  \cite{Hatta:2011gh}. 
  %The extension to higher-$p$ is straightforward.
  \vskip 1pt

  Having  identified the term responsible for  the NEC-violating radiation in the 2-point correlator \eqref{46E}     
 let us try now to remove it from  $ \langle\hskip -1mm \langle   T^{i 0}(x) D^j(y)  \rangle\hskip -1mm\rangle$\,. 
    Using
     the expression \eqref{OD}  for
     $ \langle\hskip -1mm \langle  {\cal O}(x) D^j(y) \,\rangle\hskip -1mm\rangle$, 
  with $\Delta_{\cal O}=n-2$,   gives
%\footnote{The argument works for any worldvolume coordinate $x^\beta$, not only the time coordinate $x^0$.}
 $$
- \xi \, \partial^i \partial^0 \langle\hskip -1mm \langle\,    {\cal O} D^j  \rangle\hskip -1mm\rangle\, =\,
%\hskip -38mm  \partial^i \partial^\beta \biggl[   {x^j  \vert x_\perp\vert^{p-n+2}\over \vert x\vert^{2p+2}}\biggr] \,=\,
 \xi b_{\cal O}\,(2p+2) \partial^i   \Bigl( x^0 x^j\, { \vert x_\perp\vert^{p-n+2}\over \vert x\vert^{2p+4}}\Bigr)\,=\, 
$$\vskip -6mm
\bea\label{derivs}
=\, \xi b_{\cal O}\, (2p+2)\,  { x^0  \,\vert x_\perp\vert^{p-n+2}\over \vert x\vert^{2p+4}}\Bigl( \delta^{ij}+  (p\hskip -0.4mm - \hskip -0.4mm n+2) {x^i x^j \over  \vert x_\perp\vert^{ 2}}
-  (2p+4)   { x^i x^j\over \vert x\vert^{2 }}\,
\Bigr) \ . 
\eea
Adding this to  \eqref{46E} shows that  to cancel the NEC-violating term  we must choose
\bea\label{bOone}
\xi \, b_{\cal O} (p+1)(p+2)  = \mathfrak{b}_1\ .  
\eea
To remove  the static stress, on the other hand,   we had to choose  $\xi$ as in 
eq.\,\eqref{xi}. When combined with \eqref{OD} this   reads
\bea
\xi_{\rm def} \, b_{\cal O} (p+1) = {a_T\over n }\, (p+1)\, 2^{p}\, \Gamma ({p+1\over 2} )  \pi^{-(p+1)/2} \,  =\, {\mathfrak{b}_3\over 2n}\ . 
\eea
Lo and behold,  the   two values of $\xi$  agree if and only if the conjectured  BPS relation  \eqref{conjj} is satisfied\,!
 This proves the boxed assertion  of  section \ref{sec:4.1}. 
  \vskip 1pt
  
   Adding 
  \eqref{46E} and \eqref{derivs}, and using  the value  \eqref{bOone} of  $\xi b_{\cal O}$ 
   that restores  the NEC  we find
  \bea\label{318}
\langle\hskip -1mm \langle \,\tilde T^{i0}(x)  D^j (y)\rangle\hskip -1mm\rangle = 
 -\mathfrak{b}_3 (1- {1\over n}) \, { \vert x_\perp\vert^{p-n+2}\over \vert x\min y\vert^{2p+4}}
 (x^0\min y^0)  \delta^{ij} \, +\, {\rm [non\ susy]}\  , 
   \eea
 where   [non\ susy]  
  vanishes  iff   \eqref{conjj} is obeyed.   The above  $\propto \delta^{ij}$  correlation function 
  %makes only velocity-dependent contributions to the energy flux
  is required by  relativistic 
     covariance. The reason is that the energy flux of a defect moving at constant speed in the $j$th direction can be computed  either by a Lorentz boost of the energy-momentum
      $\langle\hskip -1mm \langle \, \tilde T^{\mu\nu}\rangle\hskip -1mm\rangle $ at rest,  
     eq.\,\eqref{312}, or equivalently  by  inserting  the operator  $\exp(iv \int dy^0\, y^0  D^j)$.  Agreement of the two results  fixes uniquely the residual correlator \eqref{318}. 
  
  It would be interesting to understand whether supersymmetry allows a  proper derivation of the Schott term in  the Lorentz-Dirac equation.
Absorbing the leading singularities  of $T^{\mu\nu}$ by redefining  the mass looks at odds, as we have explained,   with relativistic invariance when the fields carry non-zero stress in the 
particle's rest frame. This is  the case  for the Coulomb field, and it could be the source of the 
encountered   difficulties in standard electrodynamics. 
  We defer the question to  future work.
  
       A last  remark concerns  the modified energy-momentum tensor. Since  dynamical defects break anyway the  conformal symmetry,  the fact  that $\tilde T^{\ \mu}_{\mu}\not= 0$ 
      is not  vexing. But the  modification \eqref{modstress}   relies on the 
       existence of an   R-singlet scalar  operator  of dimension $n-2$,  coupling appropriately to all superconformal defects. 
         Candidates  exist  in the stress-tensor  multiplet of most bulk SCFTs, see  ref.\,\cite{Cordova:2016emh}, but  with  two notable exceptions:  ${\cal N}=1$   in four  dimensions
      and  ${\cal N}=2$   in three dimensions.  
      The BPS equation \eqref{conj} continues to hold in these cases, but 
     ${\cal O}$ is just  a fictitious intermediate device of  the   proof.

 %%%%%%%%%%%%%%%%%%%
%%%%%%%%%%%%%%%%%%%%%%%%%%%%%%%%%%%%%%%%
%%%%%%%%%%%%%%%%%%%%%%%%%%%%%%%%%%%%%%%%

\newcommand{\braket}[1]{\langle\hskip -1mm \langle #1  \rangle\hskip -1mm\rangle}

\section{Proof of the conjecture}\label{sec:proof}
 Let us  come then to the   proof of the  conjecture \eqref{conjj}. 
 All we will use from the previous  section is eq.\,\eqref{318} 
 which implies,  in particular, 
 that  $\langle\hskip -1mm \langle \,\tilde T^{i0}(x)  D^j (y)\rangle\hskip -1mm\rangle =0$ for $i\not= j$. 
Pick two transverse coordinates, say $x^{p}$ and $x^{p+1}$,  and define $ z=x^p + ix^{p+1}$.
%\footnote{\,Up to now $x^{0, 1, \cdots , p-1}$ were
%coordinates along the defect worldvolume. This change of conventios  will hopefully not  cause   confusion.
%}  
We have then equivalently 
\bea\label{51}
\langle\hskip -1mm \langle \, T_{\,\alpha z}(x) 
 D_z (y)\rangle\hskip -1mm\rangle\  \propto\   {\partial\over \partial x^\alpha}  {\partial\over \partial  z} \left( {\bar z\, \vert x_\perp\vert^{p-n+2}\over \vert x\min y\vert^{2p+2}}
\right)  \quad
{\rm iff} \quad {\mathfrak{b}_1\over \mathfrak{b}_3}  = {p+2\over 2n}\ . 
\eea
We used  here the rotation symmetry to get rid of the $i$\,$=$\,$j$ components on  the left, plus  the fact that the last two terms  of \eqref{46E}
combine as in  the right-hand side above  if and only if   $\mathfrak{b}_1/\mathfrak{b}_3$ is   fixed as in  \eqref{conjj}. 
Note  that time  can be   replaced   by any  $x^\alpha$ along the defect worldvolume, and  that the only role of the 
operator  ${\cal O}$ is to determine  the  right-hand side of \eqref{51}.

   This form  of the conjecture is appealing because all   superconformal defects  
share  a common unbroken  $\mathfrak{su}(1,1|1) \supset \mathfrak{so}(2,1)\oplus \mathfrak{so}(2)_{\rm def}$
 symmetry, as one can verify from table \ref{tab1}. 
 The
 $\mathfrak{so}(2,1)$ subgroup of this   superalgebra   is the conformal group in the worldvolume time  $x^0$  (or $x^+$ for surface defects),  
 the $\mathfrak{so}(2)_{\rm def}$ rotates  the transverse $z$-plane (and in certain cases also the R-charges), 
and the two supercharges $Q, \bar{Q}$ obey  $\{Q, \bar{Q}\} $\,=\,$  P^0$ or $P^+$. 
Since $T_{z0}$ is a  top component of the full bulk superalgebra,  it is also the  top component 
of an  $\mathfrak{su}(1,1|1)$  multiplet whose  
 superprimary  ${\cal A}$ has dimension $\Delta_{\cal A} = n-1$. 
 One can then try  to 
  prove the conjecture in a single  go  from  the Ward identity $0 = \bar{Q} Q \braket{{\cal A} D_z}$, using  the fact that the displacement operator
  is also an $\mathfrak{su}(1,1|1)$  top component \cite{Agmon:2020pde}.

 The  strategy works nicely  for line defects of a three-dimensional ${\cal N}=2$ SCFT, but computing   the general form of  $\braket{{\cal A} D_z}$ turned out to be  arduous. 
We have thus followed   an alternative  route  that uses   a single, judiciously-chosen supercharge   $Q$, and the following
 slight  reformulation of  the conjecture  
 \bea\label{51b}
\langle\hskip -1mm \langle \, T_{\,z z}(x)  D_z (y)\rangle\hskip -1mm\rangle\  \propto\   {\partial^2\over \partial z^2}   \left( {\bar z\, \vert x_\perp\vert^{p-n+2}\over \vert x\min y\vert^{2p+2}}
\right)  \quad
{\rm iff} \quad {\mathfrak{b}_1\over \mathfrak{b}_3}  = {p+2\over 2n}\ . 
\eea
%This follows from   $\braket{\Tilde{T}_{zz}(x)  D_z(0)} = 0$ iff  the conjecture holds,   a fact which we  leave to the reader to verify. 
 This  can be derived in the same way as \eqref{51}. 
 %  The unbroken supersymmetry  $Q$  will be  chosen to make  the transformation  of $T_{zz}$ particularly simple.  
\vskip 2pt

  To avoid clattering, we will change completely our notation 
  in this final section,  hoping  this will not confuse the reader.
  %  In order to unify the proofs we introduce  some notation. 
   We work in Euclidean signature   and   let  the transverse coordinate $z$ be  $x_1+ix_2$,\footnote{\,So that a lower $z$-index stands for  $V_z = {1\over 2}(V_1-iV_2)$.}   and   the 
 imaginary time   be  one of $x_{3,4,5}$. 
 We also now use  early  Greek letters as spinor indices. The index structure of the gamma matrices  in $n=3,5$ dimensions is $\gamma_\mu =(\gamma_\mu)_\alpha^{\ \beta}$\,, while  in $n=4,6$ %dimensions
 it is $\gamma_\mu =(\gamma_\mu)_\alpha^{\ \dot \beta}$ and $\bar \gamma_\mu =(\gamma_\mu)_{\dot \alpha}^{\  \beta}$. 
 Indices are raised and lowered with the charge-conjugation matrices  
   $(C^{(n)})_{\alpha\beta}$ in  $n=3,5$ dimensions, $(C^{(4)})_{\alpha\beta}= (C^{(4)})_{\dot\alpha\dot\beta}$  in four dimensions,   
     and 
 $(C^{(6)})_{\alpha\dot \beta}$   in six dimensions where the conjugate spinor representations are inequivalent. 
All the bulk superalgebras of  table \ref{tab1} have   $\mathfrak{u}(1)$ and/or $\mathfrak{su}(2)$
 R-symmetries.  We denote their  generators by $r$ and   $(R_I)_{A}^{\ B}$ with $I=1,2,3$ and $A,B=1,2$. 
The index structure of the supercharges is thus  $Q_\alpha^A$ or $\bar Q_{\dot \alpha}^A$.  
 The $\mathfrak{su}(2)$ R-symmetry index is raised by multiplying with $\varepsilon^{AB}$  on the left where $\varepsilon^{12}=1$.

   In order to unify the proofs we also  choose the following  bases of 
   $\gamma$-matrices: 
     In three dimensions  we use the 2$\times$2 Pauli matrices $\sigma_i$.  In four dimensions   $\gamma_\mu = (\sigma_i , i \mathbb{I}_2)$
  and $\bar \gamma_\mu = (\sigma_i ,  - i \mathbb{I}_2)$. 
  In five dimensions  we choose the following basis\footnote{\,This basis is not the same as the one in  appendix \ref{F42}.}
\bea
\gamma_1 = \sigma_1 \otimes I_2
\,,\ \
\gamma_2 = \sigma_2 \otimes I_2
\,,\ \
\gamma_3 = \sigma_3 \otimes \sigma_1
\,,\ \
\gamma_4 = \sigma_3 \otimes \sigma_2
\,,\ \
\gamma_5 = \sigma_3 \otimes \sigma_3
\,.
\eea
In  six dimensions we add $\gamma_6 = i \mathbb{I}_4$ with  $\bar{\gamma}_6 = -i \mathbb{I}_4$ while  all
 other $(\bar{\gamma}_i)_{\dot\alpha}{}^{\beta}$ are the same as $(\gamma_i)_{\alpha}{}^{\dot\beta}$\,.
  The charge conjugation matrices are 
  $C^{(3)} = C^{(4)} = i \sigma_2$ and $C^{(5)} = C^{(6)} = \sigma_1 \otimes i \sigma_2$.
  % correct: I guess this part is not needed since it is already explained in the previous paragraph?
 % In $n=3,5$ the $\alpha$ index is lowered by multiplying $C^{(n)}$ on the right.
 % In four dimensions  both the the $\alpha$ and the $\dot\alpha$ indices are  lowered by the same matrix 
 % $(C^{(4)})_{\dot\alpha \dot\beta}= (C^{(4)})_{\alpha \beta}$.
 % In six dimensions  the $\dot{\alpha}$ index is lowered by $(C^{(6)})_{\dot\alpha \beta}$, and the $\alpha$ index is lowered by $(C^{(6)T})_{\alpha \dot\beta}$.

      Choosing the  $\gamma$-matrices as above,  and orienting   the static (hyper)planar defects in an appropriate way, 
allows us to list  the  preserved supercharges for all the DCFTs  of table \ref{tab1} in a unified way, as shown   in table \ref{tab2} below. 
All the defects are half-BPS, so there are four unbroken supercharges in the 5d and 6d  ${\cal N}=1$  and   4d
 ${\cal N}=2$ cases. The 4d  ${\cal N}=1$  and 3d  ${\cal N}=2$ DCFTs have two unbroken supercharges.
 
%HERE 

\vskip 5pt

 \definecolor{Gray}{gray}{0.92}  
   \begin{table}[tbh]
\centering

\begin{tabular}{ |c|c|c|c| } 
\hline
\rowcolor{Gray} $n$ & $p$  & defect directions &  preserved supercharges \\  
\hline \hline
\multirow{2}{3em}{\ \ \ \ 6}
& 2 & 3,4 & $Q_1^1$, $Q_3^1$, $Q_1^2$, $Q_3^2$ \\ 
& 4 & 3,4,5,6 & $Q_1^1$, $Q_2^1$, $Q_3^2$, $Q_4^2$ \\ 
\hline \hline
\multirow{3}{3em}{\ \ \ \ 5}
& 1 & 5 & $Q_1^1$, $Q_4^1$, $Q_1^2$, $Q_4^2$ \\ 
& 2 & 3,4 & $Q_1^1$, $Q_3^1$, $Q_1^2$, $Q_3^2$ \\ 
& 3 & 3,4,5 & $Q_1^1$, $Q_2^1$, $Q_3^2$, $Q_4^2$ \\ 
\hline \hline
\multirow{2}{3em}{\ \ \ \ 4} & 1 & 4 & ${\rm Re}\{\mathcal{Q}_1^1$, $\mathcal{Q}_2^1$, $\mathcal{Q}_1^2$, $\mathcal{Q}_2^2\}$ \\
& 2 & 3,4 & $Q_1$, $\bar{Q}_2$ \\
\hline \hline
3 & 1 & 3 & $Q_1$, $\bar{Q}_2$ \\
\hline
 
\end{tabular}
 \vskip 2pt
\caption{\footnotesize
The preserved supercharges for all the  DCFTs of table \ref{tab1}. The directions 1,2 are always transverse to the defect worldvolume,
while the  parallel directions are shown in the third column above. 
 {For  $(n,p)=(4,1)$, the  supercharges are complex and only  the real parts shown in the table are unbroken.}
Our proof  uses  Ward identities of the first  preserved supercharge in each case.  
%
% (for instance $Q_1^1$ or $Q_1$)  in all cases.
 }
\label{tab2}
\end{table}

We want to show that the special form of $\langle\hskip -1mm \langle \, T_{\,z z}(x)  D_z (y)\rangle\hskip -1mm\rangle\ $ in \eqref{51b} follows from superconformal
Ward identities.  To this end we need the
 transformations of various fields in the stress-tensor multiplets. We   list these transformation  in appendix \ref{app:C}. They are 
 obtained from  the Jacobi identities for the superconformal algebras which are recorded for completeness in appendix \ref{algebras}. 
The transformation  of $T_{\mu\nu}$ under the $Q$-generators has the  universal form
\bea
Q^A_{\alpha} (T_{\mu\nu}) = \scalebox{1.25}{${1\over 4}$} (\partial_\rho J^A_{\mu\beta}) (\gamma^\rho{}_\nu)_\alpha{}^\beta + (\mu \leftrightarrow \nu)
\ , 
\eea
where the index $A$ can be dropped if the R-symmetry is $\mathfrak{u}(1)$. 
We will need  a single preserved supercharge:  
 $Q_1^1$ in 6d and 5d,  Re$(Q_1^1)$ in 4d $\mathcal{N}=2$, and $Q_1$ for the last two rows of the table.
 Our previous painstaking choices  were designed to simplify the action of this preserved supercharge on  $T_{zz}$.

 A second fact that we will need is that  there exists  a fermionic defect operator $\Lambda$ in the displacement multiplet such that  $Q (\Lambda) = D_z$.
Note indeed that there is always a broken supercharge $Q_{\rm br}$ such that $\{Q, Q_{\rm br}\} = P_z$. If we choose
 $\Lambda$ to be the corresponding goldstino then   $Q(\Lambda) = D_z$ up to a derivative of a lower  component. 
We show that this derivative-term is  absent by working out explicitly the transformation of the displacement multiplets   in appendix \ref{app:C}.

We are now ready for the proof. 
For the 4d surface defects  and the 3d  line defects the result follows from  
  Ward identities of  the supercharge $Q = Q_1$.   
Its action on  $T_{zz}$ and the $\mathfrak{u}(1)$ R-symmetry current $j_z$ reads
\bea
Q (T_{zz}) = \scalebox{1.25}{${1\over 2}$} \partial_z J_{z 1}
\ ,\quad\quad
Q (j_z) = \scalebox{1.25}{${1\over 2}$} J_{z 1}
\ , 
\eea
where $J_{\mu\alpha}$ is the supercurrent. 
Then\, $Q \braket{T_{zz}(x) \Lambda(y)} = Q \braket{j_z(x) \Lambda(y)}=0$ gives
\bea
\braket{T_{zz}(x) D_z(y)} = {\partial \over \partial z}  \braket{j_z(x) D_z(y)}
\propto {\partial^2\over \partial z^2}   \left( {\bar z\, \vert x_\perp\vert^{p-n+2}\over \vert x\min y\vert^{2p+2}}
\right)
\ .
\eea
We also  used here the fact that the  kinematic structure 
of  $\braket{j_\mu(x) D_j(y)}$\,\footnote{\,This vanishes  for  codimension other than two.} is fixed as   above by conformal symmetry.

In 5d and 6d  we need  the transformation  of $T_{zz}$, $j_z^3$ and $O$ under the  preserved supercharge $Q = Q_1^1$, 
where $j_\mu^3$ is $R^{I=3}$ component of the $\mathfrak{su}(2)$ R-symmetry current, and $O$ is the scalar  in the stress-tensor multiplet. One finds
\bea
Q (T_{zz}) = \scalebox{1.25}{${1\over 2}$} \partial_z J_{z1}^1
\ ,\quad
Q (j_z^3) = \scalebox{1.25}{${1\over 2}$} J_{z1}^1 + \eta\, \partial_z \chi_1^1
\ ,\quad
Q (O) = \chi_1^1
\ .
\eea
The   factor $\eta$ is $\frac{1}{5}$ in 6d and $\frac{1}{4}$ in 5d.
Supersymmetry implies that  $Q \braket{T_{zz}(x) \Lambda(y)}$, $Q \braket{j_z^3(x) \Lambda(y)}$ and $Q \braket{O(x) \Lambda(y)}$ all vanish. Combining these Ward identities
 and using the rigid kinematic structure of  $\braket{j_z^3(x) D_z(y)}$  gives the desired result
\bea
\hskip -4mm \braket{T_{zz}(x) D_z(y)} =  {\partial \over \partial z}  \braket{j_z^3(x) D_z(y)} \min  \eta\, {\partial^2\over \partial z^2}   \braket{O(x) D_z(y)} \propto {\partial^2\over \partial z^2}   \left( {\bar z\, \vert x_\perp\vert^{p-n+2}\over \vert x\min y\vert^{2p+2}}
\right)
.\ \ 
\eea

Finally, for 4d  line defects  we need the transformation  of $T_{zz}$, $j_z^3$ and $O$ under $Q = Q_1^1 + \bar{Q}_1^1$, where $j_\mu^3$ is the $R^{I=3}$ component of the $\mathfrak{su}(2)$ R-symmetry current, and $O$ is the scalar superprimary in the stress-tensor multiplet,\footnote{
These transformations  can be also found in \cite{Fiol:2015spa, Bianchi:2019sxz}. Our $j_{\mu}^3$ is the $t_{\mu 1}{}^1$ of this reference. 
%We also record the transformation law in Appendix \ref{algebras} in our notation.
}
\bea
\hskip -4mm Q (T_{zz}) \hskip -0.3mm=\hskip -0.3mm \scalebox{1.2}{${1\over 2}$} \partial_z (J_{z1}^1 \hskip -0.3mm +\hskip -0.3mm  \bar{J}_{z1}^1)
\,,\  
Q (j_z^3) \hskip -0.3mm=\hskip -0.3mm \scalebox{1.2}{${1\over 2}$} (J_{z1}^1 
\hskip -0.3mm+ \hskip -0.3mm \bar{J}_{z1}^1) \min \scalebox{1.2}{${1\over 3}$} \partial_z (\chi_1^1 \min \bar{\chi}_1^1)
\,,\  
Q (O) \hskip -0.3mm=\hskip -0.3mm \chi_1^1 \min \bar{\chi}_1^1
\,.\ \ 
\eea
The vanishing of  $Q \braket{T_{zz}(x) \Lambda(y)} $, $Q \braket{j_z^3(x) \Lambda(y)}$ and $Q \braket{O(x) \Lambda(y)}$ then gives
\bea
\braket{T_{zz}(x) D_z(y)} = 
\frac{1}{3}  {\partial^2\over \partial z^2} 
\braket{O(x) D_z(y)}
\propto {\partial^2\over \partial z^2}   \left( {\bar z\, \vert x_\perp\vert^{p-n+2}\over \vert x\min y\vert^{2p+2}}
\right)
\,.
\eea

This  completes the proof of  (\ref{51b})  in all the cases.
   
 %%%%%%%%%%%%%%%%%%%%%%%%%%%%%%  
    \vskip 5mm 
 {\bf Aknowledgements}:  We  thank  Nadav Drukker, Luc Frappat, Marco Meineri  and Paul Sorba for discussions. We also thank the Pollica Physics Center for hosting the workshop ``Defects, from condensed matter to quantum gravity'', where this project started. 
 CB aknowledges  the hospitality of the physics department  at the National Technical University of Athens, 
   and thanks  the greek ministry of education for support   
through grant no.\,Y$\Pi$2TA-0559198. 
 %ΥΠ2ΤΑ-0559198

 \medskip
 %%%%%%%%%%%%%%%%%%%%%%%%%%%%%

\newpage
 %%%%%%%%%%%%%%%%%%%%%%%%%%%%%%%%%%%%%%%%%%%%%%%%
 
\appendix
\numberwithin{equation}{section}

\section{Real subalgebras  of $F(4;2)$}\label{F42}

The   bosonic algebra of   $F(4;2)$ that generates the symmetries of the ${\cal N}=1$ SCFT in five dimensions is 
  $\mathfrak{so}(2,5)\oplus  \mathfrak{su}(2)_R$. The fermionic   generators are  a (spinor, doublet) of this algebra, see e.g.\,\cite{FVP}. 
 They  are split   by their dilatation charge  in  raising  and lowering  operators, 
viz.  Poincar\'e supercharges $Q$ and  conformal supercharges $S$. 

   To understand their reality properties it is convenient to start with a Weyl  spinor of $\mathfrak{so}(2,8)$
which also transforms  as (spinor, doublet) of  $\mathfrak{so}(2,5)\oplus  \mathfrak{su}(2)_R$. 
The $Q$ and $S$ supercharges  are   $\mathfrak{so}(8)$ spinors with  opposite chirality. 
 Imposing the  Majorana condition  leaves  eight  real Poincar\'e and as many  conformal supercharges, which is 
 precisely the content of $F(4;2)$. In short 
 $Q \in  (\boldsymbol{8}_s, \boldsymbol{2}) $ and $S \in  (\boldsymbol{8}_c, \boldsymbol{2}) $ where $\boldsymbol{8}_s$ and
$ \boldsymbol{8}_c$   are the  two  inequivalent spinors of $\mathfrak{so}(8)$.

 Note that  from the  viewpoint of $\mathfrak{so}(2,5)\oplus  \mathfrak{su}(2)_R$, 
 the  $\mathfrak{so}(8)$  Majorana  condition  is  a {\it symplectic-Majorana}  condition.\footnote{A nice general discussion  of this condition can be found in \cite{FOF}.} 
One can understand why  using the standard  basis of the Clifford algebra  in $2k$ dimensions in terms of  Pauli matrices
 \bea
&\hskip -70mm k=1: \quad  \gamma^1= \sigma^2\ , \ \   \gamma^2= \sigma^3\ ;  \nonumber  \\
&k=2: \quad  \gamma^1= \sigma^2\otimes\sigma^1\ , \ \   \gamma^2= \sigma^3\otimes\sigma^1\ , \ \ \gamma^3=  \boldsymbol{1}\otimes \sigma^2\ ,  \nonumber \ \ 
\gamma^4=  \boldsymbol{1}\otimes \sigma^3\  ; 
 \eea
and so on till $k=4$. In this basis the Majorana condition for $\mathfrak{so}(8)$ spinors is $\Psi ^*= B\Psi$,  
 where $B= \gamma^1 \gamma^3 \gamma^5 \gamma^7$ is  the product of all  imaginary $\gamma$-matrices \,\cite{Polchinski:1998rr}. 
Writing   $\Psi$ as a  doublet of $\mathfrak{so}(5)$ spinors gives  $\Psi_a^*=  i\epsilon_{ab} (C \Psi)^b$
 where $C= i\gamma^1 \gamma^3 \gamma^5$. Note that $C C^* = -1$ which  is why $\mathfrak{so}(5)$ spinors are pseudoreal, and why there are no Weyl-Majorana 
 $\mathfrak{so}(6)$ spinors.

   Consider now the real  subalgebras of $F(4;2)$  corresponding  to superconformal  line and  surface defects. For line defects 
 the unbroken bosonic symmetry is 
\bea
%correct formula
\mathfrak{so}(2,1)\oplus \mathfrak{so}(4)_{\rm def} \oplus  \mathfrak{su}(2)_R\  \simeq
\  \mathfrak{so}(2,1)\oplus \mathfrak{so}(3)\oplus \mathfrak{so}(3)^\prime\oplus  \mathfrak{su}(2)_R \,,  
\eea
with  the 
 $\mathfrak{so}(2,8)$ spinor transforming in  the $(\boldsymbol{2},  \boldsymbol{2}, 0, \boldsymbol{2})\oplus (\boldsymbol{2}, 0,  \boldsymbol{2},  \boldsymbol{2})$
 representation.  It is clearly compatible to keep only half of this spinor by 
  imposing the  projection $\gamma^2 \gamma^3 \gamma^4 \gamma^5=+$. This gives  four real Poincar\'e and as many conformal supercharges,  which is precisely
  the fermionic content of   $D(2,1;2;0)$.  
The residual commuting $\mathfrak{su}(2)_{\rm c}$   in  eq.\,\eqref{1in5}
 is  identified with $\mathfrak{so}(3)^\prime$\,. 
 
    For surface defects,  on the other hand, the unbroken bosonic symmetry is 
 \bea
%correct formula
\mathfrak{so}(2,2)\oplus \mathfrak{so}(3)_{\rm def} \oplus  \mathfrak{su}(2)_R\  \simeq\  \mathfrak{so}(2,1)_+\oplus \mathfrak{so}(2,1)_-\oplus \mathfrak{so}(3)_{\rm def} \oplus  \mathfrak{su}(2)_R \,,  
\eea   
  and   the 
 $\mathfrak{so}(2,8)$ spinor transforms in  the $(\boldsymbol{2}, 0,  \boldsymbol{2},   \boldsymbol{2})\oplus (0, \boldsymbol{2},   \boldsymbol{2},  \boldsymbol{2})$
 representation.    The compatible projection   is now a chiral  projection on the string worldsheet, and the   commuting subalgebra is 
 %correct
  $\mathfrak{so}(2,1)_-$ or $\mathfrak{so}(2,1)_+$. 
  This agrees with eq.\,\eqref{twoin5}

   To confirm this analysis, we have also constructed explicitly the  real  superalgebra embeddings for the above line and surface defects.  
     
%%%%%%%%%%%%%%%%%%%%%%%%%%%%%%%%%%%%%%%%%%%%%%%%%%%%%
 
%%%%%%%%%%%%%%%%%%%%%%%%%%%%%%%%%%%%%%%%%%%%%%%%%%%%% 
\newcommand{\beg}{\begin{equation}\begin{gathered}{}}
\newcommand{\eeg}{\end{gathered}\end{equation}}

\section{Superconformal algebras}\label{algebras}
The common part of all superconformal algebrs is the conformal algebra $\mathfrak{so}(2,d)$,
\beg
[D, P_\mu] = P_\mu \ ,\quad\quad
[D, K_\mu] = - K_\mu \ ,\\
[M_{\mu \nu}, M_{\rho \sigma}] = \eta_{\mu \rho} M_{\nu \sigma} + \eta_{\nu\sigma} M_{\mu\rho} - \eta_{\nu\rho} M_{\mu\sigma} - \eta_{\mu \sigma} M_{\nu \rho}
\ ,\\
[M_{\mu\nu}, P_{\rho}] = \eta_{\mu\rho} P_{\nu} - \eta_{\nu\rho} P_{\mu}
\ ,\quad\quad
[M_{\mu\nu}, K_{\rho}] = \eta_{\mu\rho} K_{\nu} - \eta_{\nu\rho} K_{\mu}
\ ,\\
[P_{\mu}, K_{\nu}] = 2 M_{\mu\nu} - 2 \eta_{\mu\nu} D
\ .
\eeg
We work in the real form where $(P_\mu)^\dagger = K^{\mu}$. When there is a $\mathfrak{su}(2)$ symmetry its generators are normalized 
by $[R^I, R^J] = i \, \varepsilon^{IJK} R_K$. 

The fermionic generators $Q$ and $S$ carry a spinor and an R-symmetry index which determine their
commutators with the bosonic generators.
The $M_{\mu\nu}$ generators in the spinor representation are $\scalebox{0.99}{$\frac{1}{2}$}  (\gamma_{\mu\nu})_\alpha{}^\beta$
or $\scalebox{0.99}{$\frac{1}{2}$}   (\bar{\gamma}_{\mu\nu})_{\dot\alpha}{}^{\dot\beta}$, so for example
 $[M_{\mu\nu}, Q_{\alpha}^A] =\scalebox{0.99}{$\frac{1}{2}$}  (\gamma_{\mu\nu})_\alpha{}^\beta Q_{\beta}^A$ etc. 
The $\mathfrak{su}(2)$
R-symmetry  generators are $\scalebox{0.99}{$\frac{1}{2}$} \sigma^I$\,.
When there is also a $\mathfrak{u}(1)$   the  generators  $Q$ have 
 R-charge \scalebox{0.99}{+$\frac{1}{2}$} while  the   generators $S$ have R-charge \scalebox{0.99}{$-\frac{1}{2}$}. 
Finally we  always have
\bea
[D,Q] = \scalebox{0.99}{$\frac{1}{2}$} Q \,, \ \ 
[D,S] = - \scalebox{0.99}{$\frac{1}{2}$} S \, \ \ \  {\rm and}\ \ \ 
[P_\mu,Q] =  [K_\mu,S] = 0 \ . 
\eea

The anticommutators $\{Q,Q\}$ and $\{S,S\}$ are also   fixed up to an overall   factor. 
The  only non-trivial anticommutator is $\{Q,S\}$.
Schematically $\{Q, S\} \sim M + D + R$,\\  and one determines  the coefficient  of each term by imposing the super-Jacobi identities.
We now give the   remaining (anti-)commutators   for the  superalgebras of table \ref{tab1}. 
The classical superalgebras  can   be found in ref.\,\cite{Frappat:1996pb}, the exceptional  $F(4;2)$ in section 20.2.1 of ref.\,\cite{FVP}, 
and  all  line-defect superalgebras in ref.\,\cite{Agmon:2020pde}. 
The conventions in  these references differ however  from ours.

\vskip 5mm

%% \subsection{$\mathfrak{osp}(8^*|4)$}
 
  \leftline{  
\fbox{ $\mathfrak{osp}(8^*|4)$ 
     }}
\vskip 5pt
 
This is the superconformal algebra of 6d $\mathcal{N} = (1,0)$ with 
 $\mathfrak{su}(2)$ R-symmetry:
\beg
 [K_{\mu}, Q_{\alpha}^A] = (\gamma_\mu)_\alpha{}^{\dot\beta} S_{\dot\beta}^A
\ ,\quad\quad
[P_{\mu}, S_{\dot\alpha}^A] = - (\bar{\gamma}_\mu)_{\dot\alpha}{}^\beta Q_{\beta}^A
\ ,\\
\{Q_{\alpha}^A, Q_{\beta}^B\} = 2\, \varepsilon^{AB} (\gamma^\mu)_{\alpha\beta} P_{\mu}
\ ,\quad\quad
\{S_{\dot\alpha}^{A}, S_{\dot\beta}^B\} = -2 \, \varepsilon^{AB} (\bar{\gamma}^\mu)_{\dot\alpha \dot\beta} K_{\mu}
\ ,\\
\{Q_{\alpha}^A, S_{\dot\beta}^B\} =
\varepsilon^{AB} (\gamma^{\mu\nu})_{\alpha \dot\beta} M_{\mu\nu}
- 2\, \varepsilon^{AB} (C^{(6)T})_{\alpha \dot\beta} D
- 8 \, (\sigma_I)^{AB} (C^{(6)T})_{\alpha \dot\beta}  R^I
\ .
\eeg

\vskip 2mm

  \leftline{  
\fbox{ $F(4;2)$ 
     }}
\vskip 5pt
 
This is the superconformal algebra of 5d $\mathcal{N} = 1$ with 
 $\mathfrak{su}(2)$ R-symmetry:
  \beg
 [K_{\mu}, Q_{\alpha}^A] = (\gamma_\mu)_\alpha{}^\beta S_{\beta}^A
\ ,\quad\quad
[P_{\mu}, S_{\alpha}^{A}] = - (\gamma_\mu)_\alpha{}^\beta Q_{\beta}^A
\ ,\\
\{Q_{\alpha}^A, Q_{\beta}^B\} = 2\, \varepsilon^{AB} (\gamma^\mu)_{\alpha\beta} P_{\mu}
\ ,\quad\quad
\{S_{\alpha}^A, S_{\beta}^B\} = 2\, \varepsilon^{AB} (\gamma^\mu)_{\alpha\beta} K_{\mu}
\ ,\\
\{Q_{\alpha}^A, S_{\beta}^B\} = 
-\varepsilon^{AB} (\gamma^{\mu\nu})_{\alpha\beta} M_{\mu\nu}
+ 2\, \varepsilon^{AB} (C^{(5)})_{\alpha\beta} D
+ 6\, (\sigma_I)^{AB} (C^{(5)})_{\alpha\beta} R^I
\ .
\eeg

\vskip 2mm

  \leftline{  
\fbox{ $\mathfrak{su}(2,2|2)$ 
     }}
\vskip 5pt

% \subsection{$\mathfrak{su}(2,2|2)$}
  This is the superconformal algebra of 4d $\mathcal{N} = 2$ with  R-symmetry $\mathfrak{su}(2)\oplus \mathfrak{u}(1)$:
 \beg
 [K_{\mu}, Q_{\alpha}^A] =  (\gamma_{\mu})_{\alpha}{}^{\dot\beta} \bar{S}_{\dot\beta}^A
\ ,\quad\quad
[K_{\mu}, \bar{Q}_{\dot\alpha}^A] =(\bar{\gamma}_{\mu})_{\dot\alpha}{}^{\beta} S_{\beta}^A
\ ,\\
[P_{\mu}, S_{\alpha}^A] = - (\gamma_{\mu})_{\alpha}{}^{\beta} \bar{Q}_{\beta}^A
\ ,\quad\quad
[P_{\mu}, \bar{S}_{\dot\alpha}^A] = - (\bar{\gamma}_{\mu})_{\dot\alpha}{}^{\beta} Q_{\beta}^A
\ ,\\
\{Q_{\alpha}^A, \bar{Q}_{\dot\beta}^B\} = 2 \, \varepsilon^{AB} (\gamma^{\mu})_{\alpha \dot\beta} P_{\mu}
\ ,\quad\quad
\{S_{\alpha}^A, \bar{S}_{\dot\beta}^B\} = 2 \, \varepsilon^{AB} (\gamma^{\mu})_{\alpha \dot\beta} K_{\mu}
\ ,\\
\{Q_{\alpha}^A, S_{\beta}^B\} = \varepsilon^{AB} (\gamma^{\mu \nu})_{\alpha \beta} M_{\mu \nu}
- 2\, \varepsilon^{AB} (C^{(4)})_{\alpha \beta} (D-r)
- 4\, (\sigma_I)^{AB} (C^{(4)})_{\alpha \beta} R^I
\ ,\\
\{\bar{Q}_{\dot\alpha}^A, \bar{S}_{\dot\beta}^B\} = -\varepsilon^{AB} (\bar{\gamma}^{\mu \nu})_{\dot\alpha \dot\beta} M_{\mu \nu}
+ 2\, \varepsilon^{AB}(C^{(4)})_{\dot\alpha \dot\beta} (D+r)
+ 4\, (\sigma_I)^{AB} (C^{(4)})_{\dot\alpha \dot\beta} R^I
\ .
\eeg

\vskip 2mm

  \leftline{  
\fbox{ $\mathfrak{su}(2,2|1)$ 
     }}
\vskip 5pt

% \subsection{$\mathfrak{su}(2,2|1)$}
 This is the superconformal algebra of 4d $\mathcal{N} = 1$ with 
 $\mathfrak{u}(1)$ R-symmetry:
\beg
 [K_{\mu}, Q_{\alpha}] =  (\gamma_{\mu})_{\alpha}{}^{\dot\beta} \bar{S}_{\dot\beta}
\ ,\quad\quad
[K_{\mu}, \bar{Q}_{\dot\alpha}] =(\bar{\gamma}_{\mu})_{\dot\alpha}{}^{\beta} S_{\beta}
\ ,\\
[P_{\mu}, S_{\alpha}] = - (\gamma_{\mu})_{\alpha}{}^{\beta} \bar{Q}_{\beta}
\ ,\quad\quad
[P_{\mu}, \bar{S}_{\dot\alpha}] = - (\bar{\gamma}_{\mu})_{\dot\alpha}{}^{\beta} Q_{\beta}
\ ,\\
\{Q_{\alpha}, \bar{Q}_{\dot\beta}\} = 2 \, (\gamma^{\mu})_{\alpha \dot\beta} P_{\mu}
\ ,\quad\quad
\{S_{\alpha}, \bar{S}_{\dot\beta}\} = -2 \, (\gamma^{\mu})_{\alpha \dot\beta} K_{\mu}
\ ,\\
\{Q_{\alpha}, S_{\beta}\} = (\gamma^{\mu \nu})_{\alpha \beta} M_{\mu \nu}
-2 \, (C^{(4)})_{\alpha \beta} (D-3\,r)
\ ,\\
\{\bar{Q}_{\dot\alpha}, \bar{S}_{\dot\beta}\} = (\bar{\gamma}^{\mu \nu})_{\dot\alpha \dot\beta} M_{\mu \nu}
- 2 \, (C^{(4)})_{\dot\alpha \dot\beta} (D+3\,r)
\ .
\eeg

\vskip 2mm

  \leftline{  
\fbox{ $\mathfrak{osp}(2 | 4; \mathbb{R})$ 
     }}
\vskip 5pt
 
% \subsection{$\mathfrak{osp}(2 | 4; \mathbb{R})$}
 
This is the superconformal algebra of 3d $\mathcal{N}=2$ with 
 $\mathfrak{u}(1)$ R-symmetry:
\beg
  [K_{\mu}, Q_{\alpha}] =  (\gamma_{\mu})_{\alpha}{}^{\beta} \bar{S}_{\beta}
\ ,\quad\quad
[K_{\mu}, \bar{Q}_{\alpha}] = (\gamma_{\mu})_{\alpha}{}^{\beta} S_{\beta}
\ ,\\
[P_{\mu}, S_{\alpha}] = - (\gamma_{\mu})_{\alpha}{}^{\beta} \bar{Q}_{\beta}
\ ,\quad\quad
[P_{\mu}, \bar{S}_{\alpha}] = - (\gamma_{\mu})_{\alpha}{}^{\beta} Q_{\beta}
\ ,\\
\{Q_{\alpha}, \bar{Q}_{\beta}\} = 2 \,(\gamma^{\mu})_{\alpha \beta} P_{\mu}
\ ,\quad\quad
\{ S_{\alpha}, \bar{S}_{\beta}\} = -2 \, (\gamma^{\mu})_{\alpha \beta} K_{\mu}
\ ,\\
\{ Q_{\alpha}, S_{\beta}\} = (\gamma^{\mu \nu})_{\alpha \beta} M_{\mu \nu}
- 2 \, (C^{(3)})_{\alpha \beta} (D-2\,r)
\ ,\\
\{\bar{Q}_{\alpha}, \bar{S}_{\beta}\} = (\gamma^{\mu \nu})_{\alpha \beta} M_{\mu \nu}
- 2 \, (C^{(3)})_{\alpha \beta} (D+2\,r)
\ .
\eeg

\vskip 2mm

  \leftline{  
\fbox{ $D(2,1;\lambda,0)$ 
     }}
\vskip 5pt

% \subsection{$D(2,1;\lambda,0)$}
 This is a 1d superconformal algebra with four supercharges.
The R-symmetry algebra is $\mathfrak{su}(2)_r \times \mathfrak{su}(2)_R$. The  respective generators  $r^I$ and $R^I$. 
are normalized as above. 
We use $a,b=1,2$ for  the $\mathfrak{su}(2)_r$ index, and $A,B=1,2$ for  the $\mathfrak{su}(2)_R$ index. 
The (anti-)commutators read
\beg
 [K, Q^{Aa}] = S^{Aa}
\ ,\quad\quad
[P, S^{Aa}] = -Q^{Aa}
\ ,\\
\{Q^{Aa}, Q^{Bb}\} = 2 \, \varepsilon^{AB} \varepsilon^{ab} P
\ ,\quad\quad
\{S^{Aa}, S^{Bb}\} = 2 \, \varepsilon^{AB} \varepsilon^{ab} K
\ ,\\
\{Q^{Aa}, S^{Bb}\} =
- 2\lambda \, \varepsilon^{AB}(\sigma_I)^{ab} r^I
+ 2 \, \varepsilon^{AB} \varepsilon^{ab} D
+ 2(\lambda+1) \, (\sigma_I)^{AB} \varepsilon^{ab} R^I
\ .
\eeg
The special case $\lambda=1$ is isomorphic to $\mathfrak{osp}(4^*|2)$,  and the case $\lambda=0$ is isomorphic to $\mathfrak{psu}(1,1|2)\oplus \mathfrak{su}(2)$
which contains $\mathfrak{su}(1,1|1)$ as a subalgebra.

%%%%%%%%%%%%%%%

\section{Supersymmetry transformations}\label{app:C}
In this appendix we present  the transformation under the action of the Poincar\'e supercharges $Q$ of the stress-tensor and displacement multiplets.

The stress tensor multiplets for all bulk  superconformal algebras in $n\geq 3$ dimensions can be found in \cite{Cordova:2016emh}.
We restrict  to  the minimal SCFTs of  table \ref{tab1}.
 The four-dimensional case  was worked out,  in a slightly different notation,   in \cite{Fiol:2015spa, Bianchi:2019sxz}.
For the six-dimensional   $\mathcal{N}=(2,0)$ SCFT  the $Q$-actions  can be found in \cite{Trepanier:2021xmr}.
We used the technique of  this latter reference to
work out all  other cases.

  To be more precise, let $Q(A)$ be the action of $Q$ on some  operator  $A$.
 We begin with  the most general ansatz for $Q(A)$ that is consistent with the field content of the multiplet and
the bosonic symmetries.
We also  impose  the conservation and zero-trace  conditions of $A$,  if any.
Finally we fix the coefficients  for each tensor structure by requiring that  $\{Q,Q\} = 2\, P$, that is
 $\{Q,Q\}(A) = 2\, \partial A$.

\vskip 5mm

  \leftline{
\fbox{ $\mathfrak{osp}(8^* | 4)$
     }}
\vskip 4pt

The $(40+40)$ stress tensor multiplet contains a scalar operator $O$, a spinor $\chi_{\alpha}^A$, a self-dual 3-form $H_{[\mu\nu\rho]}$, the $\mathfrak{su}(2)$ R-symmetry currents
 $j_{\mu}^I$, the supersymmetry currents  $J_{\mu \alpha}^A$ and the stress tensor $T_{\mu\nu}$.\footnote{The self-duality condition
  is $(\,^*H)^{\mu\nu\rho} = i H^{\mu\nu\rho}$ and
the supercurrents satisfy $(\bar{\gamma}^\mu)_{\dot\alpha}{}^\beta J_{\mu\beta}^M = 0$.
%All the currents $j_\mu^I$, $J_{\mu\alpha}^A$ and $T_{\mu \nu}$ are conserved, and $T_{\mu \nu}$ is of course traceless.
}
Schematically
\bea
O
\overset{Q}{\longrightarrow}
\chi_{\alpha}^A
\overset{Q}{\longrightarrow}
j_{\mu}^I \oplus H_{\mu\nu\rho}
\overset{Q}{\longrightarrow}
J_{\mu \alpha}^A
\overset{Q}{\longrightarrow}
T_{\mu\nu}
\ .
\eea
The explicit transformations are
\beg
Q_{\alpha}^A (O) = \chi_{\alpha}^A
\ ,\\
Q_{\alpha}^A (\chi_{\beta}^B) =
j_{\mu}^I \,(\gamma^\mu)_{\alpha\beta} (\sigma_I)^{AB}
+ H_{\mu\nu\rho} \,(\gamma^{\mu\nu\rho})_{\alpha\beta} \,\varepsilon^{AB}
+ (\partial_{\mu} O) \,(\gamma^\mu)_{\alpha\beta} \,\varepsilon^{AB}
\ , \\
Q_{\alpha}^A (j_{\mu}^I) = \scalebox{1.25}{${1\over 2}$}
 J_{\mu\alpha}^B \, (\sigma^I)_B{}^A
- \scalebox{1.25}{${1\over 5}$} (\partial_{\nu} \chi_{\beta}^B) \, (\sigma^I)_B{}^A (\gamma_{\mu}{}^{\nu})_\alpha{}^\beta
\ ,\\
Q_{\alpha}^A (H_{\mu\nu\rho}) = - \scalebox{1.25}{${1\over 48}$} J_{[\mu |\beta|}^A \, (\gamma_{\nu\rho]})_\alpha{}^\beta
+ \scalebox{1.25}{${1\over 30}$} (\partial_{\sigma} \chi_{\beta}^A) \, (\gamma^\sigma \bar{\gamma}_{\mu\nu\rho})_\alpha{}^\beta
\ ,\\
Q_{\alpha}^A (J_{\mu\beta}^B) = 2\, T_{\mu\nu} \, (\gamma^\nu)_{\alpha\beta} \,\varepsilon^{AB}
%\\ \quad\quad
- \scalebox{1.25}{${2\over 5}$}(\partial_{\nu} j_{\rho}^I) \, (\gamma_{\mu}{}^{\nu\rho} - 4\, \delta_{\mu}{}^\rho \gamma^\nu)_{\alpha\beta} \,\varepsilon^{AC} (\sigma_I)_C{}^B \\
 \quad\quad
+ \scalebox{1.25}{${2\over 5}$} (\partial_{\nu} H_{\rho\sigma\lambda}) \, (\delta_\mu{}^\nu \gamma^{\rho\sigma\lambda}
- 6\, \delta_{\mu}{}^\rho \gamma^{\nu\sigma\lambda}
- 18\, \delta_{\mu}{}^{\rho} \eta^{\nu\sigma} \gamma^{\lambda})_{\alpha\beta} \,\varepsilon^{AB}
\ ,\\
Q_{\alpha}^A (T_{\mu\nu}) = \scalebox{1.25}{${1\over 4}$} (\partial_{\rho} J_{\mu \beta}^A) \, (\gamma^{\rho}{}_{\nu})_{\alpha}{}^{\beta} + (\mu \leftrightarrow \nu)
\ .
\eeg

\vskip 2mm

  \leftline{
\fbox{ $F(4;2)$
     }}
\vskip 5pt

The $(32+32)$ stress tensor multiplet contains a scalar $O$, a spinor $\chi_{\alpha}^A$, a 2-form $B_{\mu\nu}$, the $\mathfrak{su}(2)$ R-symmetry currents $j_{\mu}^I$, the supercurrents
 $J_{\mu \alpha}^A$ and
  $T_{\mu\nu}$,
 \bea
O
\overset{Q}{\longrightarrow}
\chi_{\alpha}^A
\overset{Q}{\longrightarrow}
j_{\mu}^I \oplus B_{\mu\nu}
\overset{Q}{\longrightarrow}
J_{\mu \alpha}^A
\overset{Q}{\longrightarrow}
T_{\mu\nu}
\ .
 \eea
The explicit transformations are
 \beg
Q_{\alpha}^A (O) = \chi_{\alpha}^A
\ ,\\
Q_{\alpha}^A (\chi_{\beta}^B) =
j_{\mu}^I \,(\gamma^\mu)_{\alpha\beta} (\sigma_I)^{AB}
+ B_{\mu\nu} \,(\gamma^{\mu\nu})_{\alpha\beta} \,\varepsilon^{AB}
 + (\partial_{\mu} O) \,(\gamma^\mu)_{\alpha\beta} \,\varepsilon^{AB}
\ ,\\
Q_{\alpha}^A (j_{\mu}^I) = \scalebox{1.25}{${1\over 2}$} J_{\mu\alpha}^B \, (\sigma^I)_B{}^A
- \scalebox{1.25}{${1\over 4}$}(\partial_{\nu} \chi_{\beta}^B) \, (\sigma^I)_B{}^A (\gamma_{\mu}{}^{\nu})_\alpha{}^\beta
\ ,\\
Q_{\alpha}^A (B_{\mu\nu}) = - \scalebox{1.25}{${1\over 4}$} J_{\rho \beta}^A \, (\gamma_{\mu \nu}{}^{\rho})_{\alpha}{}^{\beta}
+\scalebox{1.25}{${1\over 8}$} (\partial_{\rho} \chi_{\beta}^A) \, (2 \, \gamma_{\mu \nu}{}^{\rho} + \delta_{[\mu}{}^{\rho} \gamma_{\nu]})_\alpha{}^\beta
\ ,\\
Q_{\alpha}^A (J_{\mu\beta}^B) = 2\, T_{\mu\nu} \, (\gamma^\nu)_{\alpha\beta} \,\varepsilon^{AB}
%\\ \quad\quad
- \scalebox{1.25}{${1\over 2}$} (\partial_{\nu} j_{\rho}^I) \, (\gamma_{\mu}{}^{\nu\rho} - 3\, \delta_{\mu}{}^\rho \gamma^\nu)_{\alpha\beta} (\sigma_I)^{AB} \\
 \quad\quad
% correct: the sign before $\gamma_{\mu \nu \rho \sigma}$
+ \scalebox{1.25}{${1\over 2}$} (\partial_{\nu} B_{\rho \sigma}) \, (\gamma_{\mu}{}^{\nu \rho \sigma}
- 2\, \delta_{\mu}{}^{\rho} \gamma^{\nu \sigma}
+ 2\, \eta^{\nu \rho} \gamma_{\mu}{}^{\sigma}
- 6\, \delta_{\mu}{}^{\rho} \eta^{\nu \sigma}
)_{\alpha\beta} \,\varepsilon^{AB}
\ ,\\
Q_{\alpha}^A (T_{\mu\nu}) = \scalebox{1.25}{${1\over 4}$}  (\partial_{\rho}J_{\mu\beta}^A) \, (\gamma^{\rho}{}_{\nu})_{\alpha}{}^{\beta} + (\mu \leftrightarrow \nu)
\ .
 \eeg

\vskip 2mm

  \leftline{
\fbox{ $\mathfrak{su}(2,2|2)$
     }}
\vskip 5pt

The $(24+24)$ stress tensor multiplet contains a scalar $O$, a spinor $\chi_\alpha^A$ (and $\bar{\chi}_{\dot\alpha}^A$), a symmetric bispinor $H_{(\alpha \beta)}$ (and $\bar{H}_{(\dot\alpha \dot\beta)}$),
the $\mathfrak{u}(1)$
and $\mathfrak{su}(2)$ R-symmetry
currents  $j_\mu$ and  $j_\mu^I$, the supercurrent $J_{\mu \alpha}^A$ (and $\bar{J}_{\mu \dot\alpha}^A$) and   $T_{\mu\nu}$,
\bea
O
\overset{Q}{\longrightarrow}
\chi_{\alpha}^A \oplus \bar{\chi}_{\dot\alpha}^A
\overset{Q}{\longrightarrow}
H_{\alpha \beta}
\oplus j_{\mu} \oplus j_{\mu}^I \oplus
\bar{H}_{\dot\alpha \dot\beta}
\overset{Q}{\longrightarrow}
J_{\mu \alpha}^A \oplus \bar{J}_{\mu \dot\alpha}^A
\overset{Q}{\longrightarrow}
T_{\mu\nu}
\ .
\eea
The explicit transformations are
\beg
Q_{\alpha}^A (O) = \chi_{\alpha}^A
\ ,\quad
\bar{Q}_{\dot\alpha}^A (O) = -\chi_{\dot\alpha}^A
\ ,\\
Q_{\alpha}^A (\chi_{\beta}^B) = H_{\alpha \beta} \,\varepsilon^{AB}
\ ,\quad
\bar{Q}_{\dot\alpha}^A (\bar{\chi}_{\dot\beta}^B) = \bar{H}_{\dot\alpha \dot\beta} \,\varepsilon^{AB}
\ ,\\
\bar{Q}_{\dot\alpha}^A (\chi_{\beta}^B) =
[- j_\mu  \varepsilon^{AB} + j_\mu^I (\sigma_I)^{AB}
- (\partial_\mu O) \varepsilon^{AB} ] (\bar{\gamma}^\mu)_{\dot\alpha \beta}
\ ,\\
Q_{\alpha}^A (\bar{\chi}_{\dot\beta}^B) =
[j_\mu  \varepsilon^{AB}
+ j_\mu^I (\sigma_I)^{AB}
- (\partial_\mu O) \varepsilon^{AB}] \,
(\gamma^\mu)_{\alpha \dot\beta}
\ ,\\
\bar{Q}_{\dot\alpha}^A (H_{\beta \gamma}) =
[\scalebox{1.25}{${1\over 2}$}J_{\mu \beta}^A
+\scalebox{1.25}{${2\over 3}$} \partial_\mu \chi_\beta^A] \,
(\bar{\gamma}^\mu)_{\dot\alpha \gamma} + (\beta \leftrightarrow \gamma)
\ ,\\
\ \
Q_\alpha^A (\bar{H}_{\dot\beta \dot\gamma}) =
[\scalebox{1.25}{${1\over 2}$} \bar{J}_{\mu \dot\beta}^A -\scalebox{1.25}{${2\over 3}$}\partial_\mu\bar{\chi}_{\dot\beta}^A]
(\gamma^\mu)_{\alpha \dot\gamma} +
(\dot\beta \leftrightarrow \dot\gamma)
\ ,\\
Q_{\alpha}^A (j_\mu ) = \scalebox{1.25}{${1\over 2}$}J_{\mu \alpha}^A - \scalebox{1.25}{${2\over 3}$} (\partial_\nu \chi_{\beta}^A) (\gamma_{\mu}{}^{\nu})_{\alpha}{}^{\beta}
\ ,%\\
\ \
\bar{Q}_{\dot\alpha}^A (j_\mu) = -\scalebox{1.25}{${1\over 2}$} \bar{J}_{\mu \dot\alpha}^A - \scalebox{1.25}{${2\over 3}$}
 (\partial_\nu \bar{\chi}_{\dot\beta}^A) (\bar{\gamma}_{\mu}{}^{\nu})_{\dot\alpha}{}^{\dot\beta}
\ ,\\
Q_{\alpha}^A (j_\mu^I) = [\scalebox{1.25}{${1\over 2}$} J_{\mu \alpha}^B + \scalebox{1.25}{${1\over 3}$}(\partial_\nu \chi_{\beta}^B) \,(\gamma_\mu{}^\nu)_{\alpha}{}^{\beta}] \,(\sigma^I)_B{}^A
\ , \\
\
\bar{Q}_{\dot\alpha}^A (j_\mu^I) = [\scalebox{1.25}{${1\over 2}$} \bar{J}_{\mu \dot\alpha}^B -
\scalebox{1.25}{${1\over 3}$} (\partial_\nu \bar{\chi}_{\dot\beta}^B) \, (\bar{\gamma}_{\mu}{}^{\nu})_{\dot\alpha}{}^{\dot\beta}] \, (\sigma^I)_B{}^A
\ ,\\
Q_{\alpha}^A (J_{\mu \beta}^B) = (\partial_\nu H_{\gamma\beta}) \, \varepsilon^{AB} (\gamma_{\mu}{}^{\nu})_{\alpha}{}^{\gamma} + \scalebox{1.25}{${1\over 3}$} (\partial_\nu H_{\gamma\alpha}) \, \varepsilon^{AB} (\gamma_{\mu}{}^{\nu})_{\beta \gamma}
\ ,\\
\bar{Q}_{\dot\alpha}^A (\bar{J}_{\mu \dot\beta}^B) =
(\partial_\nu \bar{H}_{\dot\gamma \dot\beta}) \, \varepsilon^{AB} (\bar{\gamma}_\mu{}^\nu)_{\dot\alpha}{}^{\dot\gamma}
-\scalebox{1.25}{${1\over 3}$}(\partial_\nu \bar{H}_{\dot\gamma \dot\alpha}) \, \varepsilon^{AB} (\bar{\gamma}_{\mu}{}^{\nu})_{\dot\beta \dot\gamma}
\ ,\\
\bar{Q}_{\dot\alpha}^A (J_{\mu\beta}^B) = -2\, T_{\mu\nu} \,\varepsilon^{AB} (\bar{\gamma}^\nu)_{\dot\alpha \beta} + \scalebox{1.25}{${1\over 3}$} [(\partial_\nu j_\rho) \, \varepsilon^{AB} + 2 (\partial_\nu j_\rho^I) \, (\sigma_I)^{AB}] \, (\bar{\gamma}_{\mu}{}^{\nu\rho} - 2\, \delta_\mu{}^\rho \bar{\gamma}_{\nu})_{\dot\alpha \beta}
\ ,\\
Q_{\alpha}^A (\bar{J}_{\mu \dot\beta}^B) = 2\, T_{\mu\nu} \,\varepsilon^{AB} (\gamma^\nu)_{\alpha \dot\beta} + \scalebox{1.25}{${1\over 3}$} [(\partial_\nu j_\rho) \, \varepsilon^{AB} - 2 (\partial_\nu j_\rho^I) \, (\sigma_I)^{AB}] \,
(\gamma_\mu{}^{\nu\rho} - 2\, \delta_\mu{}^\rho \gamma^\nu)_{\alpha \dot\beta}
\ ,\\
Q_{\alpha}^A (T_{\mu \nu}) = \scalebox{1.25}{${1\over 4}$} (\partial_\rho J_{\mu \beta}^A) \, (\gamma^\rho{}_\nu)_{\alpha}{}^\beta + (\mu \leftrightarrow \nu)
\ , \\
\  \bar{Q}_{\dot\alpha}^A (T_{\mu \nu}) = \scalebox{1.25}{${1\over 4}$} (\partial_\rho \bar{J}_{\mu \dot\beta}^A) \, (\bar{\gamma}^\rho{}_\nu)_{\dot\alpha}{}^{\dot\beta} + (\mu \leftrightarrow \nu)\ .
\eeg

\vskip 2mm

  \leftline{
\fbox{ $\mathfrak{su}(2,2|1)$
     }}
\vskip 5pt

 This is a subalgebra of the previous case, but we give  it  separately  for the reader's convenience.
 The $(8+8)$ stress-tensor multiplet contains the $\mathfrak{u}(1)$ R-symmetry current $j_{\mu}$, the supercurrents $J_{\mu \alpha},\bar{J}_{\mu \dot\alpha}$ and   $T_{\mu\nu}$,
 \bea
j_{\mu}
\overset{Q}{\longrightarrow}
J_{\mu \alpha} \oplus \bar{J}_{\mu \dot\alpha}
\overset{Q}{\longrightarrow}
T_{\mu\nu}
\ .
 \eea
 Notice that this multiplet contains  no scalar, the super-primary is the R-symmetry current.
The explicit  transformations are
 \beg
Q_{\alpha} (j_\mu) = \scalebox{1.25}{${1\over 2}$} J_{\mu \alpha}
\ ,\quad\quad
\bar{Q}_{\alpha} (j_\mu) = - \scalebox{1.25}{${1\over 2}$} \bar{J}_{\mu \alpha}
\ ,\\
\bar{Q}_{\dot\alpha} (J_{\mu \beta}) = 2\, (\bar{\gamma}^\nu)_{\dot\alpha \beta} T_{\mu \nu}
- (\partial_{\nu} j_{\rho})\, ( \bar{\gamma}_{\mu}{}^{\nu\rho} - 2\, \delta_{\mu}{}^{\rho} \bar{\gamma}^\nu)_{\dot\alpha \beta}
\ ,\\
Q_{\alpha} (\bar{J}_{\mu \dot\beta}) = 2\, (\gamma^\nu)_{\alpha \dot\beta} T_{\mu \nu}
+ (\partial_{\nu} j_{\rho})\, ( \gamma_{\mu}{}^{\nu\rho} - 2\, \delta_{\mu}{}^{\rho} \gamma^\nu)_{\alpha \dot\beta}
\ ,\\
Q_{\alpha} (T_{\mu \nu}) = \scalebox{1.25}{${1\over 4}$} (\partial_{\rho} J_{\mu \beta})\, (\gamma^{\rho}{}_{\nu})_{\alpha}{}^{\beta}  + (\mu \leftrightarrow \nu)
\ ,\\
\bar{Q}_{\dot\alpha} (T_{\mu \nu}) = \scalebox{1.25}{${1\over 4}$} (\partial_{\rho} \bar{J}_{\mu \dot\beta})\, (\bar{\gamma}^{\rho}{}_{\nu})_{\dot\alpha}{}^{\dot\beta} + (\mu \leftrightarrow \nu)
\ .
 \eeg

\vskip 2mm

  \leftline{
\fbox{ $\mathfrak{osp}(2 | 4;\mathbb{R})$
     }}
\vskip 5pt

The $(4+4)$ stress-tensor multiplet contains the $\mathfrak{u}(1) $ R-symmetry current $j_{\mu}$, the supercurrents  $J_{\mu \alpha},\bar{J}_{\mu \alpha}$ and   $T_{\mu\nu}$,
 \bea
j_{\mu}
\overset{Q}{\longrightarrow}
J_{\mu \alpha} \oplus \bar{J}_{\mu\alpha}
\overset{Q}{\longrightarrow}
T_{\mu\nu}
\ .
 \eea
The explicit transformations are
\beg
Q_{\alpha} (j_\mu) = \scalebox{1.25}{${1\over 2}$}J_{\mu \alpha}
\ ,\quad\quad
\bar{Q}_{\alpha} (j_\mu) = -\scalebox{1.25}{${1\over 2}$}\bar{J}_{\mu \alpha}
\ ,\\
\bar{Q}_{\alpha} (J_{\mu \beta}) = 2\, (\gamma^\nu)_{\alpha \beta} T_{\mu \nu}
- 2\, (\partial_{\nu} j_{\rho})\, (\gamma_{\mu}{}^{\nu\rho} - \delta_{\mu}{}^{\rho} \gamma^\nu)_{\alpha \beta}
\ ,\\
Q_{\alpha} (\bar{J}_{\mu \beta}) = 2\, (\gamma^\nu)_{\alpha \beta} T_{\mu \nu}
+ 2\, (\partial_{\nu} j_{\rho})\,(\gamma_{\mu}{}^{\nu\rho} - \delta_{\mu}{}^{\rho} \gamma^\nu)_{\alpha \beta}
\ ,\\
Q_{\alpha} (T_{\mu \nu}) = \scalebox{1.25}{${1\over 4}$}(\partial_{\rho} J_{\mu \beta})\, (\gamma^{\rho}{}_{\nu})_{\alpha}{}^{\beta}  + (\mu \leftrightarrow \nu)
\ ,\\
\bar{Q}_{\alpha} (T_{\mu \nu}) =\scalebox{1.25}{${1\over 4}$} (\partial_{\rho} \bar{J}_{\mu \beta})\, (\gamma^{\rho}{}_{\nu})_{\alpha}{}^{\beta} + (\mu \leftrightarrow \nu)
\ .
\eeg

\vskip 4mm

The  transformation  of the displacement multiplets under the preserved  supercharges is  much simpler.
The generic multiplet contains the displacement vector  $D$, a  fermion $\Lambda$,  and in some of the cases a scalar $\Phi$.
 This is  the structure under the transverse-rotation group $\mathfrak{so}(n\min p)_{\rm def}$.  The fermion and the preserved supercharges
$Q$ are also worldvolume spinors, and  they may carry an extra $r$-symmetry index $A$.
  The defect superalgebras of table \ref{tab1} were recorded in appendix \ref{algebras}.
The transformation laws are fixed by   imposing the super-Jacobi identities.

We  label the cases by $(n,p)$ and group them according to the codimension and the number of preserved supercharges.
What is required for our  proof is that there always exists a supercharge $Q$ and a fermion  $\Lambda$ such that $Q(\Lambda) = D_z$.

\vskip 2mm
  \leftline{
\fbox{ $(4,2)$ and $(3,1)$
     }}
\vskip 5pt
This is the simplest multiplet, 
 generated by  two  supercharges and without a scalar  component $\Phi$. Complexifying  the transverse $\mathfrak{so}(2)_{\rm def}$ index
gives
  \beg
\{Q,\Lambda\} = D
\ ,\quad
[\bar{Q}, D] = 2\, \partial \Lambda\ ,
\
\eeg
and  the  conjugate relations.
Here $\partial$ stands for $\partial_t$ when $p=1$ and for $\partial_+$ when $p=2$.

\vskip 3mm
  \leftline{
\fbox{ $(6,4)$ and $(5,3)$
     }}
\vskip 5pt

%The displacement multiplet contains a scalars $\Phi$, a spinor $\Lambda_\alpha$ and the displacement operators $D$ (together with their conjugates $\bar{\Phi}, \bar{\Lambda}_{\dot\alpha}, \bar{D}$).
 For $(n,p)=(6,4)$
 the non-vanishing (anti-)commutators are
 \beg
[Q_\alpha, \Phi] = \Lambda_\alpha
\ ,\quad
[\bar{Q}_{\dot\alpha}, D] = -2\, (\bar\gamma^i)_{\dot\alpha}{}^{\beta} \partial_i \Lambda_\beta
\ ,\\
\{Q_{\alpha}, \Lambda_{\beta}\} = C^{(p)}_{\alpha \beta} D
\ ,\quad
\{ \bar{Q}_{\dot\alpha}, \Lambda_{\beta}\} = 2\, (\bar{\gamma}^i)_{\dot\alpha \beta} (\partial_i \Phi)
\ .
\eeg
The  index $i$ runs  from $1$ to $p$, and we have again complexified the transverse $\mathfrak{so}(2)_{\rm def}$ indices.
For $(n,p)=(5,3)$ one  must replace  $\bar{\gamma}^i \to \gamma^i$ and $\dot\alpha \to \alpha$.

\vskip 3mm
  \leftline{
\fbox{ $(6,2)$ and $(5,1)$
     }}
\vskip 5pt

Here $Q$ and $\Lambda$ are complex $\mathfrak{so}(4)$ spinors  and they have an additional $r$-symmetry index.
The   displacement can be written as  a bispinor $D^{a\dot b}$ and there is no scalar $\Phi$.
The non-vanishing (anti-)commutators read
\beg
\{Q^{Aa},   \Lambda^{B\dot b}\} = \varepsilon^{AB} D^{a\dot b}
\ ,\quad
[Q^{Aa}, D^{b\dot c}] = -2\, \varepsilon^{ab}\, \partial \Lambda^{A\dot c}
\ ,
\eeg
together with  the  conjugate relations.

\vskip 3mm
  \leftline{
\fbox{ $(5,2)$ and $(4,1)$
     }}
\vskip 5pt

Here $Q$ and $\Lambda$ are transverse-$\mathfrak{so}(3)$ spinors  and have an additional $r$-symmetry index.
% correct: antisymmetric -> symmetric, because we raise the index b using $\varepsilon$ and the matrices are symmetric
The   displacement   is a symmetric  bispinor $D^{ab}=(\sigma_j)^{ab} D^j$ ($j=1,2,3)$.
The non-vanishing (anti-)commutators read
\beg
[Q^{Aa}, \Phi] = \Lambda^{Aa}
\ ,\quad
[Q^{Aa}, D^{bc}] = -(\varepsilon^{ac}\, \partial \Lambda^{Ab}  + \varepsilon^{ab}\, \partial \Lambda^{Ac})
\ ,\\
\{Q^{Aa}, \Lambda^{Bb}\} = \varepsilon^{AB} (D^{ab} + \varepsilon^{ab} \partial \Phi)
\ ,
\eeg
This is the only tricky case because $\{Q, \Lambda\} \sim D + \partial \Phi$.
But with the choices made in  section \ref{sec:proof} one  finds  $D_z \propto D^{22}$,
and the $\partial \Phi$ term drops out in the anticommutator  $\{Q^{12}, \Lambda^{22}\}$\,.
This is sufficient for our proof.
%%%%%%%%%%%%%%%%%%%%%%%%%%%%%%%%%%%%%%%%%%%%%%%%%%%%%

 %%%%%%%%%%%%%%%%%%%%%%%%%%%%%%%%%%%%%%%%%%%%%%%%%%%%%%%

\end{document}